\documentclass[aps,prl,twocolumn,superscriptaddress,longbibliography,floatfix]{revtex4-2}

\usepackage{amsmath,amssymb,mathtools,bm}
\usepackage{graphicx}
\usepackage{xcolor}
\usepackage{tikz}
\usepackage{hyperref}
\hypersetup{colorlinks=true, linkcolor=blue!60!black, citecolor=blue!60!black, urlcolor=blue!60!black}
\providecommand{\Tr}{\mathrm{Tr}}

\providecommand{\ket}[1]{|#1\rangle}
\providecommand{\bra}[1]{\langle #1|}

\begin{document}
\title{Quantized topological invariant of symmetry-projected Gibbs states}

\author{Weiguang Cao}
\email{weiguang@ust.hk}
\affiliation{Department of Physics, Hong Kong University of Science and Technology, Clear Water Bay, Hong Kong, China}
\affiliation{Center for Theoretical Condensed Matter Physics, Hong Kong University of Science and Technology, Clear Water Bay, Hong Kong, China}

\author{Haruki Watanabe}
\email{hwatanabe@ust.hk}
\affiliation{Department of Physics, Hong Kong University of Science and Technology, Clear Water Bay, Hong Kong, China}
\affiliation{Center for Theoretical Condensed Matter Physics, Hong Kong University of Science and Technology, Clear Water Bay, Hong Kong, China}
\affiliation{Institute for Advanced Study, Hong Kong University of Science and Technology, Clear Water Bay, Hong Kong, China}

\date{\today}

\begin{abstract}
We study Gibbs states projected onto the symmetric sector of contractible
one-form symmetries. In a three-dimensional cluster-model interpolation, this
projection stabilizes symmetry-protected topological and
projected-paramagnetic phases, both sharply distinct from thermal disorder. A
flux-twisted membrane invariant distinguishes these three phases by the values
$-1,+1,0$, respectively. These values are exact on the endpoint, self-dual,
zero-temperature, and infinite-temperature lines; elsewhere their
quantization requires positive spatial-sheet, winding-line, and interface
tensions. Quantum Monte Carlo supports the quantization through tension
diagnostics and direct finite-size estimates.
\end{abstract}

\maketitle

\textit{Introduction.}---At zero temperature, a symmetry-protected topological
(SPT) Hamiltonian cannot be deformed into a trivial one without breaking the
protecting symmetry or closing the bulk
gap~\cite{PhysRevB.80.155131,Pollmann:2009ryx,Pollmann:2009mhk,Chen:2010zpc,PhysRevB.84.235128,Chen:2011pg}.
Finite-temperature Gibbs ensembles generally lose this distinction:
fixed-point group-cohomology models are isospectral to trivial paramagnets, and
onsite symmetry does not protect against thermal defects~\cite{RYKB}.
Projection can instead impose strong symmetry: for a generator $S$, a state of
fixed charge $e^{i\theta}$ obeys $S\rho=e^{i\theta}\rho$, rather than only
$S\rho S^\dagger=\rho$~\cite{Ma:2022pvq,deGroot:2021vdi,Lessa:2024gcw,LBC};
such states can retain entanglement at arbitrarily high
temperatures~\cite{Negari:2025jdc}. Can projection also create a
finite-temperature transition when the thermal free energy before projection is
analytic?

The three-dimensional Raussendorf--Bravyi--Harrington (RBH) cluster
model~\cite{RBH} is a
canonical test. Its protecting one-form symmetry has closed-loop defects, and
Ref.~\cite{RYKB} showed that thermal SPT order survives when the locally
generated contractible charges are projected while noncontractible charges are
thermally summed. Contractible constraints admit local energetic penalties,
whereas noncontractible charges label global winding sectors. This establishes
the finite-temperature SPT endpoint, but not the interpolation to an otherwise
free trivial paramagnet.

\begin{figure}[t]
\centering
\includegraphics[width=\columnwidth]{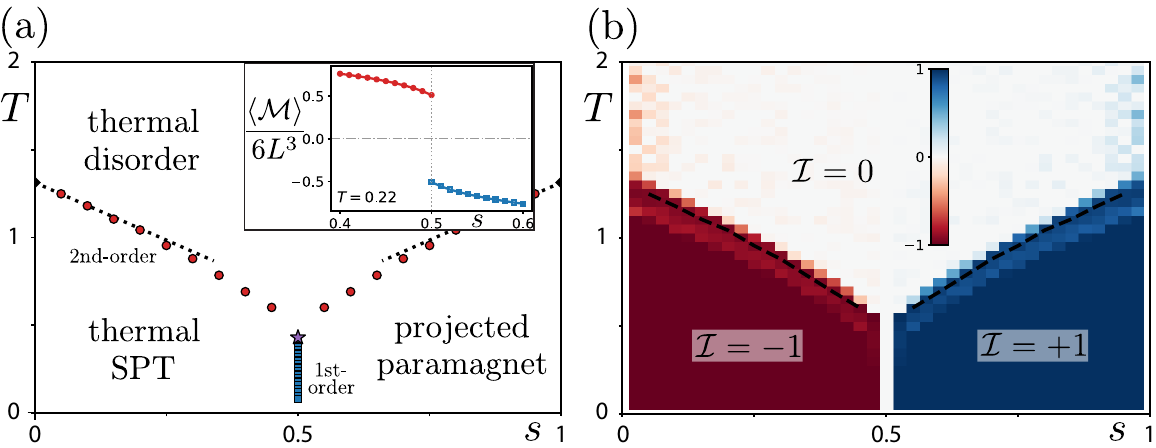}
\caption{(a)~Finite-regulator estimates of the projected phase diagram.
Red circles, blue squares, black diamonds, and the star denote the
$\beta\mu=4$ arms, first-order wall, exact endpoints, and wall endpoint;
red bars are finite-size-scaling fit errors, and black dotted lines are the
endpoint tangents in Eq.~(\ref{eq:endpoint-tangents}). Inset: stable-branch
$\langle\mathcal M\rangle/(6L^3)$ data at $L=12$, $\beta\mu=4$, and $T=0.22$,
where $\mathcal M\coloneqq\partial_sH$; red circles and
blue squares denote deconfined- and confined-prepared runs. (b)~The $L=8$
conditional finite-size QMC proxy $I_8$ for $\mathcal I$, obtained from a
positive worldsheet-sector
splitting, conditional on the spatial-sheet suppression condition in
Eq.~(\ref{eq:H-letter}). Red, blue, and white denote $-1,+1,0$; black dashed
curves are the finite-regulator arm estimates.}
\label{fig:maps}
\end{figure}

In this Letter, we show that projection creates such a transition and map
the resulting phase diagram. At the microscopically free paramagnetic endpoint,
each projected species maps
exactly to the three-dimensional Ising model~\cite{Wegner}, yielding
low-temperature projected-paramagnetic and high-temperature thermally
disordered phases. The finite-depth cluster entangler preserves the
contractible projector and reflects the interpolation, carrying the transition
to the SPT endpoint. Finite-regulator quantum Monte Carlo (QMC) finds two mirror
continuous-transition arms and brackets the endpoint of a self-dual
first-order line at the midpoint (Fig.~\ref{fig:maps}). Thus projection
stabilizes both thermal SPT and projected-paramagnetic regions, whose phase
structure belongs to the projected ensemble rather than the Hamiltonian alone.

Free-energy singularities locate the boundaries, not the topological character
of the three regions. We propose a flux-twisted membrane index $\mathcal I$
that distinguishes them and is fixed unconditionally at the endpoints by the
exact Ising mapping. In the interior, an exact identity expresses $\mathcal I$ through a
winding-sector free-energy splitting. On the SPT side, positive winding-line
and interface tensions select one sector, provided purely spatial
noncontractible sheets are exponentially suppressed. Reflection then fixes the
projected-paramagnetic side. QMC measures the sector splitting and a positive
spatial-sheet cost. A separate finite-size
calculation retaining all spatial-sheet sectors directly reconstructs the
physical index at representative low- and high-temperature points. The
resulting $L=4$ values directly support the predicted phase values without the
spatial-sheet assumption.

\textit{Setup.}---We set $k_B=1$ and measure energy in the unit coupling of
the endpoint Hamiltonians. We put one qubit on every link $l$ and plaquette $p$ of an
$L\times L\times L$ cubic $3$-torus. Let $X_l,Z_l$ and $X_p,Z_p$ denote
Pauli operators on the link and plaquette qubits, respectively, and define
$B_l\coloneqq\prod_{p\ni l}Z_p$ and
$B_p\coloneqq\prod_{l\in\partial p}Z_l$. The cluster
Hamiltonian~\cite{RBH} and the trivial paramagnet are
\begin{equation}
\begin{aligned}
H_{\rm SPT}&\coloneqq-\textstyle\sum_lX_lB_l-\sum_pX_pB_p,\\[-2pt]
H_{\rm para}&\coloneqq-\textstyle\sum_lX_l-\sum_pX_p .
\end{aligned}
\end{equation}
For $0\le s\le1$, we interpolate between them as
\begin{equation}
H(s)\coloneqq(1-s)H_{\rm SPT}+s\,H_{\rm para}.
\end{equation}
Thus $s=0$ is the cluster-SPT endpoint and $s=1$ is the
trivial-paramagnetic endpoint.

For each cube $c$ and vertex $v$, define the local one-form symmetry
generators
\begin{equation}
S^{(1)}_c\coloneqq\prod_{p\in\partial c}X_p,\qquad
S^{(1)}_v\coloneqq\prod_{l\ni v}X_l .
\end{equation}
Here $\partial c$ denotes the plaquettes bounding cube $c$, while $l\ni v$
restricts the product to the links in the vertex star at $v$. These
operators generate the contractible
$\mathbb Z_2^{(1)}\times\mathbb Z_2^{(1)}$ symmetry group
\begin{equation}
G_{\rm loc}\coloneqq\big\langle\{S^{(1)}_c\}_c,\{S^{(1)}_v\}_v\big\rangle .
\end{equation}
Here $\langle\cdots\rangle$ denotes the group generated by the listed
operators. Every term of $H(s)$ commutes with $G_{\rm loc}$.

For each qubit species choose one noncontractible plane normal to each torus
direction, and denote the resulting six symmetry operators by
$S_a^{\rm nc}$, $a=1,\ldots,6$. The full one-form symmetry group is
\begin{equation}
G_{\rm full}\coloneqq\big\langle
G_{\rm loc},S^{\rm nc}_1,\ldots,S^{\rm nc}_6\big\rangle .
\end{equation}
Two parallel representatives of the same plane class multiply to a
contractible element. Hence every finite-range $G_{\rm loc}$-invariant
Hamiltonian is $G_{\rm full}$ invariant once $L$ exceeds its interaction
range: for a local Pauli string $P$ in the Hamiltonian and any
$S_a^{\rm nc}$, choose a parallel representative $\widehat S_a^{\rm nc}$
disjoint from $P$. Then
$S_a^{\rm nc}\widehat S_a^{\rm nc}\in G_{\rm loc}$ and
$[P,\widehat S_a^{\rm nc}]=0$ imply $[P,S_a^{\rm nc}]=0$.

Although $H(s)$ is $G_{\rm full}$ symmetric, a projected ensemble need not
fix every $G_{\rm full}$ charge. The two ensembles considered below use the
projectors
\begin{equation}
\begin{aligned}
P_{\rm loc}&\coloneqq\prod_c\tfrac12\big(1+S^{(1)}_c\big)
\prod_v\tfrac12\big(1+S^{(1)}_v\big),\\[-2pt]
P_+&\coloneqq P_{\rm loc}\prod_{a=1}^{6}\tfrac12\big(1+S_a^{\rm nc}\big).
\end{aligned}
\end{equation}
At inverse temperature $\beta\coloneqq1/T$, define
\begin{equation}\label{eq:Pplus-rhoplus}
\rho_{\rm loc}(s)\coloneqq\frac{P_{\rm loc}e^{-\beta H(s)}}
{\Tr[P_{\rm loc}e^{-\beta H(s)}]},\quad
\rho_+(s)\coloneqq\frac{P_+e^{-\beta H(s)}}
{\Tr[P_+e^{-\beta H(s)}]} .
\end{equation}
For every $g\in G_{\rm loc}$, $g\rho_{\rm loc}=\rho_{\rm loc}$, so
$\rho_{\rm loc}$ is strongly $G_{\rm loc}$ symmetric. Noncontractible
generators instead act only weakly:
$S_a^{\rm nc}\rho_{\rm loc}(S_a^{\rm nc})^\dagger=\rho_{\rm loc}$, because
$P_{\rm loc}$ sums rather than fixes their charges. By contrast, $P_+$ fixes
all six charges to $+1$, making $\rho_+$ strongly $G_{\rm full}$ symmetric.
Thus the ensembles differ by summed versus fixed noncontractible charges, not
Hamiltonian symmetry. The fusion projector and sector decomposition are
derived in Secs.~\ref{supp:sm:setup} and \ref{supp:sm:exact} of the Supplemental
Material (SM).

Local energetic terms can impose the contractible projection:
$\rho_{\rm loc}(s)$ is the $\mu\to\infty$ Gibbs limit of
\begin{equation}\label{eq:soft}
H(s)-\mu\Big(\sum_vS^{(1)}_v+\sum_cS^{(1)}_c\Big),
\end{equation}
whose $s=0$ case is the construction of Ref.~\cite{RYKB}. At the finite $\mu$
used in QMC, constraint violations retain thermal weight; the simulated state
only approximates $\rho_{\rm loc}$.

For an incident link--plaquette pair
$\langle l,p\rangle_\partial$, $l\in\partial p$, define
\begin{equation}
U\coloneqq\prod_{\langle l,p\rangle_\partial}CZ_{l,p}.
\end{equation}
where $CZ_{l,p}$ is the controlled-$Z$ gate on those qubits. It obeys $U^2=I$
and
\begin{equation}
UX_lU^\dagger=X_lB_l,\quad UX_pU^\dagger=X_pB_p.
\end{equation}
so $UH(s)U^\dagger=H(1-s)$. The $B$ dressings cancel pairwise in every cube,
vertex-star, and noncontractible-plane generator; hence $U$ commutes with
$P_{\rm loc}$ and preserves each noncontractible charge sector. Cyclicity gives
the exact reflection
\begin{equation}\label{eq:mirror-letter}
Z_{\rm loc}(s,T)\coloneqq\Tr[P_{\rm loc}e^{-\beta H(s)}]
=Z_{\rm loc}(1-s,T)
\end{equation}
at every $T$, separately in each sector. This equivalence does not trivialize
the SPT distinction: although the complete $U$ commutes with $G_{\rm loc}$,
its individual controlled-$Z$ gates do not, so $U$ is not a gatewise
one-form-symmetric circuit~\cite{RYKB}.

\textit{Endpoint Ising mapping.}---At the paramagnetic endpoint, projection
produces a thermodynamic singularity although the unprojected qubits are free.
Writing $\Tr_l$ for the link-qubit trace, the identical plaquette trace on the
dual cubic lattice gives
$Z_{\rm loc}(1,T)=[Z_P(\beta)]^2$, where
\begin{equation}\label{eq:ZP-def}
Z_P(\beta)\coloneqq\Tr_l\!\biggl[
\prod_v\tfrac12(1+S_v^{(1)})e^{\beta\sum_lX_l}\biggr].
\end{equation}
Let $\mathcal V$ be the vertex set, with $N_v\coloneqq L^3$ vertices and
$N_l\coloneqq3L^3$ links. Expanding the vertex-projector factor gives
$\prod_v\tfrac12(1+S_v^{(1)})=2^{-N_v}\sum_{S\subseteq\mathcal V}
\prod_{v\in S}S_v^{(1)}$.
For $S\subseteq\mathcal V$, let $B(S)$ count links with exactly one endpoint in
$S$; the single-link traces give
\begin{equation}
Z_P(\beta)=2^{-N_v}(2\cosh\beta)^{N_l}
\sum_{S\subseteq\mathcal V}(\tanh\beta)^{B(S)} .
\end{equation}
With $\sigma_v\coloneqq-1$ for $v\in S$ and $+1$ otherwise,
$B(S)=\sum_{\langle vv'\rangle}(1-\sigma_v\sigma_{v'})/2$, where
$\langle vv'\rangle$ runs over nearest-neighbor vertex pairs,
\begin{equation}\label{eq:s1-ising-letter}
Z_P(\beta)=c_\beta\sum_{\{\sigma\}}
e^{K\sum_{\langle vv'\rangle}\sigma_v\sigma_{v'}},
\quad e^{-2K}\coloneqq\tanh\beta ,
\end{equation}
where
$c_\beta\coloneqq2^{-N_v}(2\cosh\beta)^{N_l}
(\tanh\beta)^{N_l/2}$ is analytic for $0<\beta<\infty$.
Equivalently, in the $X$ basis the projector enforces even degree at every
vertex, giving
\begin{equation}
Z_P(\beta)=e^{\beta N_l}\sum_{\mathcal C\ {\rm closed}}
e^{-2\beta|\mathcal C|}.
\end{equation}
Here $|\mathcal C|$ counts links in the closed electric string configuration;
the sum includes all noncontractible winding sectors. The cube projector gives
an identical Ising copy on the dual cubic lattice.
Hence the projected paramagnet has exactly the singularities of the
three-dimensional Ising model, whose critical coupling is
$K\approx0.2216$~\cite{FerrenbergXuLandau};
Eq.~(\ref{eq:mirror-letter}) gives the same singularity at the SPT endpoint,
\begin{equation}\label{eq:TI}
T_I\coloneqq T_c^{\rm proj}(0)=T_c^{\rm proj}(1)\approx1.3133 .
\end{equation}
Because $K$ decreases with $\beta$, the Ising-ordered regime is the
\emph{high}-temperature side of the projected-paramagnet transition, where closed
electric strings proliferate. Without $P_{\rm loc}$ the qubits decouple and
the transition disappears. The same temperature is the Wegner deconfinement
temperature of the three-dimensional $\mathbb Z_2$ gauge
theory~\cite{Wegner,Hastings:2011iig,Castelnovo:2008jyy};
extended endpoint and winding-sector derivations are in
Sec.~\ref{supp:sm:exact} of the SM.

The same mapping sends the link-species soft constraint at $s=1$ to the
Ising model in Eq.~(\ref{eq:s1-ising-letter}), now in a uniform field:
\begin{equation}\label{eq:soft-ising-letter}
\Tr_l e^{\beta(\sum_lX_l+\mu\sum_vS_v^{(1)})}
=c_{\beta,\mu}\sum_{\sigma}
e^{K\sum_{\langle vv'\rangle}\sigma_v\sigma_{v'}
  +h\sum_v\sigma_v}.
\end{equation}
Here $K$ is given in Eq.~(\ref{eq:s1-ising-letter}),
$c_{\beta,\mu}$ is analytic, and
$h\coloneqq-\tfrac12\ln\tanh(\beta\mu)$. The plaquette species gives the same field
on the dual lattice. This field rounds the singularity at finite $\mu$ and
vanishes as $\mu\to\infty$.

\textit{Projected phase diagram.}---Three inputs organize the phase diagram.
First, the endpoint mapping above fixes the common endpoint temperature
$T_I$ in Eq.~(\ref{eq:TI}).
Second, the Kennedy--Tasaki (KT) image with all six noncontractible charges
fixed to $+1$, derived in
Sec.~\ref{supp:sm:exact} of the SM, contains the gauge model
\begin{equation}\label{eq:gauge-frame}
\begin{aligned}
H_g(s)&\coloneqq-(1-s)\sum_pB_p-s\sum_lX_l\\[-2pt]
&=(1-s)\bigl(-\sum_pB_p-h^x\sum_lX_l\bigr),
\quad h^x\coloneqq\frac{s}{1-s},
\end{aligned}
\end{equation}
on the direct lattice and its cell-dual counterpart on the dual lattice.
Fixing these six sectors changes $\ln Z$ only by $O(L)$, so its bulk
free-energy density and that of the $P_{\rm loc}$ ensemble differ by
$O(L^{-2})$; the explicit bound is proved below. Conjugation by
$\prod_lZ_l$ reverses $h^x$; hence the
unit-coupling critical temperature is even in $h^x$. Assuming its regular
expansion about $h^x=0$, reflection then gives
\begin{equation}\label{eq:endpoint-tangents}
T_c^{\rm proj}(s)=
\begin{cases}
T_I(1-s)+O(s^2),&s\to0,\\
T_I s+O((1-s)^2),&s\to1,
\end{cases}
\end{equation}
with endpoint slopes $-T_I$ and $+T_I$~\cite{Companion}. Third, the self-dual
transition of $H_g$ at $(s,T)=(\tfrac12,0)$ is weakly first
order~\cite{ReissSchmidt}. Its two gapped branches survive weak heating, and
Eq.~(\ref{eq:mirror-letter}) pins their coexistence to $s=\tfrac12$.

QMC provides a finite-regulator map of this analytic skeleton
(Fig.~\ref{fig:maps}(a)): two mirror
continuous-transition arms and a self-dual first-order line delimit the
thermal SPT, projected-paramagnetic, and thermally disordered phases. The
wall endpoint is estimated as $(\tfrac12,0.43(1))$. The inferred
phase-boundary topology connects the arms to this endpoint, although the
closest arm data do not resolve the junction itself. At the fixed
free-paramagnet endpoint $s=1$, varying temperature alone crosses between
the latter two phases.

We implement $P_{\rm loc}$ through Eq.~(\ref{eq:soft}) using discrete-time
worldline QMC at $L=6,8,10,12,16$ and fixed $\beta\mu=4$
($h\simeq3.4\times10^{-4}$). Finite $\mu$ rounds the continuous arms. After
subtracting the soft-penalty contribution, we fit physical-energy inflections
as $T_{\rm inf}^{(\mu)}(s,L)-T_{\rm arm}^{(\mu)}(s)\propto L^{-1/\nu}$, where
$T_{\rm inf}^{(\mu)}$ is the finite-$L$ inflection temperature,
$T_{\rm arm}^{(\mu)}$ its thermodynamic estimate at fixed $\mu$, and
$\nu=0.630$ the three-dimensional Ising correlation-length exponent. The red
points in Fig.~\ref{fig:maps}(a) use sharper energy-fluctuation peaks; this is
not a heat capacity along fixed $\beta\mu$, but after endpoint anchoring the two
estimates agree within $0.016$. We then impose mirror symmetry. Exact
projection requires $\mu\to\infty$ before $L\to\infty$, or
$h(L)L^{y_h}\to0$, with the three-dimensional Ising magnetic exponent
$y_h=2.48185$~\cite{KPSDV}. The plotted coordinates retain finite-$\mu$ and
finite-$\Delta\tau$ systematics, where $\Delta\tau$ is the imaginary-time
step; these exceed the displayed fit errors. Regulator checks are in
Sec.~\ref{supp:sm:numerics} of the SM.

The first-order line at $s=\tfrac12$ is diagnosed by
\begin{equation}
\mathcal M\coloneqq\partial_sH
=\sum_l(X_lB_l-X_l)+\sum_p(X_pB_p-X_p),
\end{equation}
for which
$U\mathcal M U^\dagger=-\mathcal M$; Hellmann--Feynman identifies its jump
with the free-energy slope. Two stable branches persist through $T=0.42$; an
$L=16$ control remains two-branched at $0.43$, whereas none is stable at
$0.44$. Independently, fixed-$\mu=4$ crossings of
$L^{3-y_h}\langle\mathcal M\rangle/(6L^3)$ for $L=6$--$12$, with the self-dual
field detuned by $L^{-y_h}$, give the same finite-regulator wall-endpoint
estimate $0.43(1)$. Because the regulator schedules differ, the crossing is
only a consistency check; both protocols are detailed in
Sec.~\ref{supp:sm:numerics} of the SM.

At $s=1$, dilute electric strings exponentially suppress nontrivial winding
sectors below $T_c^{\rm proj}$; after condensation their relative weights
equalize. Replacing $(1+S_v^{(1)})/2$ by $(1-S_v^{(1)})/2$ at two vertices
inserts open-string endpoints and maps exactly to
$\langle\sigma_v\sigma_{v'}\rangle$, which decays for bound endpoints at low
temperature and tends to a constant after high-temperature
deconfinement~\cite{Wegner}. Thus Ising order is a disorder operator of the
projected paramagnet, while no local one-form order parameter selects either
phase. This motivates the twisted nonlocal index below; the loop-gas
derivation is in Sec.~\ref{supp:sm:index-theory} of the SM.

\begin{figure}[t]
\centering
\begin{tikzpicture}[scale=0.76,>=stealth]
  \draw[gray,dashed] (0,0) -- (0.84,0.48) -- (3.84,0.48);
  \draw[gray,dashed] (0.84,0.48) -- (0.84,3.08);
  \fill[red!14] (1.62,0.24) -- (3.42,0.24) -- (3.42,2.84) -- (1.62,2.84) -- cycle;
  \draw[red!60] (1.62,0.24) -- (3.42,0.24) (3.42,2.84) -- (1.62,2.84);
  \draw[red!60,dashed] (3.42,0.24) -- (3.42,2.84);
  \fill[blue!13,fill opacity=0.85] (0,1.3) -- (3,1.3) -- (3.84,1.78) -- (0.84,1.78) -- cycle;
  \draw[blue!70] (0,1.3) -- (3,1.3) -- (3.84,1.78) -- (0.84,1.78) -- cycle;
  \draw[red!50,dotted] (1.62,1.54) -- (3.42,1.54);
  \draw[red,very thick,->] (1.62,0.24) -- (1.62,2.84);
  \draw[red,very thick,dotted] (1.62,2.84) -- (1.62,3.1);
  \draw[red,very thick,dotted] (1.62,-0.02) -- (1.62,0.24);
  \fill[red] (1.62,1.54) circle (1.7pt);
  \draw[gray] (0,0) -- (3,0) -- (3,2.6) -- (0,2.6) -- cycle;
  \draw[gray] (3,0) -- (3.84,0.48) -- (3.84,3.08) -- (3,2.6);
  \draw[gray] (0,2.6) -- (0.84,3.08) -- (3.84,3.08);
  \node[red] at (1.4,2.66) {$\gamma$};
  \node[blue!60!black,anchor=west] at (3.9,1.82) {$\Sigma$};
  \fill[red!30] (1.350,0.644) -- (1.730,0.644) -- (1.890,0.736) -- (1.510,0.736) -- cycle;
  \draw[red!80!black,thin] (1.350,0.644) -- (1.730,0.644) -- (1.890,0.736) -- (1.510,0.736) -- cycle;
  \fill[red!30] (1.350,1.494) -- (1.730,1.494) -- (1.890,1.586) -- (1.510,1.586) -- cycle;
  \draw[red!80!black,thin] (1.350,1.494) -- (1.730,1.494) -- (1.890,1.586) -- (1.510,1.586) -- cycle;
  \draw[red,very thick] (1.62,0.60) -- (1.62,0.80);
  \draw[red,very thick] (1.62,1.44) -- (1.62,1.64);
  \fill[red] (1.62,1.54) circle (1.7pt);
  \node[red!80!black,anchor=east,font=\footnotesize] at (1.28,0.63) {$B_p\to-B_p$};
  \node[red!70!black,font=\footnotesize] at (2.9,0.5) {$Z_l\to-Z_l$};
  \node[anchor=east,font=\footnotesize] at (1.52,1.72) {$\Sigma\cdot\gamma=1$};
\end{tikzpicture}
\caption{Twisted-membrane geometry. The flat cocycle lies on the plaquettes
pierced by $\gamma$ (red); cutting turns it into the link-sign sheet. The
noncontractible membrane $\Sigma$ (blue) has $\Sigma\cdot\gamma=1$ and detects
$(-1)^{\Sigma\cdot\gamma}$.}
\label{fig:twist-schematic}
\end{figure}
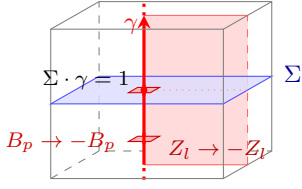

\textit{Topological index.}---We thread a one-form flux along a
noncontractible dual loop $\gamma$ with the flat, non-exact cocycle
$\eta_p\coloneqq-1$ on the pierced stack and $+1$ elsewhere.
\begin{equation}\label{eq:Htw-letter}
\begin{split}
H_{\rm tw}(s)\coloneqq{}&-(1-s)\bigl(\textstyle\sum_l X_lB_l
  +\sum_p\eta_p X_pB_p\bigr)\\
&-s\bigl(\textstyle\sum_l X_l+\sum_p X_p\bigr).
\end{split}
\end{equation}
This implements a topological twist; in a cut-open fundamental domain it
appears as a sheet of link-sign flips. On a noncontractible $2$-cycle
$\Sigma$ with odd intersection $\Sigma\cdot\gamma$, the membrane
$S_\Sigma^{(1)}\coloneqq\prod_{p\in\Sigma}X_p$ is itself a noncontractible
one-form symmetry operator. Projection with $P_+$ would make it trivial:
$S_\Sigma^{(1)}P_+=P_+$ fixes its expectation value to $+1$ for every
$s$, $T$, and $L$. By contrast, $P_{\rm loc}$ leaves both noncontractible
charge sectors in the trace, allowing the twist to polarize them:
\begin{equation}
I_L(s,T)\coloneqq\langle S_\Sigma^{(1)}\rangle_{\rm tw}^{P_{\rm loc}},\qquad
\mathcal I(s,T)\coloneqq\lim_{L\to\infty}I_L(s,T),
\end{equation}
where $\langle O\rangle_{\rm tw}^{P_{\rm loc}}\coloneqq
\Tr[O\,P_{\rm loc}e^{-\beta H_{\rm tw}(s)}]/
\Tr[P_{\rm loc}e^{-\beta H_{\rm tw}(s)}]$. The limit follows a sequence of
$L^3$ tori with fixed primitive classes $[\Sigma]$ and $[\gamma]$, and phase
values below are one-sided limits inside open phases. This is the higher-form
analogue of reading a projective endpoint charge after threading a
protecting-symmetry flux~\cite{PT,Companion};
Fig.~\ref{fig:twist-schematic} summarizes the geometry.

The index depends only on the two noncontractible classes. Moving $\Sigma$
within its class multiplies $S_\Sigma^{(1)}$ by contractible cube generators,
which act as $+1$ inside $P_{\rm loc}$; moving $\gamma$ changes $\eta_p$ by
an exact cocycle implemented by a local unitary change of variables. Within the
sector mechanism below, local symmetry-preserving perturbations vary the
relevant tensions continuously but not the intersection sign. Provided the
sector identity persists, a quantized phase value can change only when at least
one tension vanishes.

The index also obeys an exact reflection relation. Let
$W\coloneqq\prod_{p:\eta_p=-1}Z_p$; flatness gives $[W,P_{\rm loc}]=0$.
Conjugation by $U$ transfers $\eta_p$ from the $X_pB_p$ terms to the $X_p$
terms, and conjugation by $W$ completes the reflection:
$(WU)H_{\rm tw}(s)(WU)^\dagger=H_{\rm tw}(1-s)$ and
$(WU)S_\Sigma^{(1)}(WU)^\dagger=(-1)^{\Sigma\cdot\gamma}
S_\Sigma^{(1)}$. Since $WU$ preserves $P_{\rm loc}$, cyclicity gives
\begin{equation}\label{eq:twisted-reflection-letter}
I_L(s,T)=(-1)^{\Sigma\cdot\gamma}I_L(1-s,T).
\end{equation}
For odd intersection, exact finite-$L$ identities give
$I_L(\tfrac12,T)=0$, $I_L(s,0)=-1$ ($+1$) for
$s<\tfrac12$ ($s>\tfrac12$), and $I_L(s,\infty)=0$. Moreover,
$\mathcal I(0,T)=-1$ below $T_c^{\rm proj}(0)$ and $0$ above it. Proofs are in
Sec.~\ref{supp:sm:index-theory} of the SM.
Below the wall endpoint the self-dual zero belongs to coexistence rather than
either phase; above it, zero is the thermally disordered value.

\textit{Exact sector identity and quantization mechanism.}---We work in the joint
eigenbasis of the mutually commuting stabilizers $K_p\coloneqq X_pB_p$ and
$K_l\coloneqq X_lB_l$. This basis is valid at every $s$, although $H(s)$ is not
generally diagonal in it. Since
$\prod_{p\in\Sigma}B_p=I$, the membrane is diagonal,
$S_\Sigma^{(1)}=\prod_{p\in\Sigma}K_p$. The cube constraint makes the
$K_p=-1$ plaquettes a closed dual-lattice loop configuration $\mathcal C^*$.
Writing $w(\mathcal C^*)\in H_1(\mathbb T^3,\mathbb Z_2)$ for its winding
homology class, the membrane eigenvalue is the mod-$2$ pairing
$(-1)^{w(\mathcal C^*)\cdot[\Sigma]}$; each local flip changes an even number
of plaquettes of the closed $\Sigma$, leaving this parity invariant. Conjugating
the twist by the same $W$ gives
$WS_\Sigma^{(1)}W^\dagger=(-1)^{\Sigma\cdot\gamma}S_\Sigma^{(1)}$ and
$WH_{\rm tw}(s)W^\dagger=H(s)+2s\sum_{p:\eta_p=-1}X_p$. We decompose this conjugated
two-species problem into the even and odd sectors
\begin{equation}
Z_\pm\coloneqq\Tr\!\biggl[
P_{\rm loc}\frac{1\pm S_\Sigma^{(1)}}{2}
e^{-\beta WH_{\rm tw}(s)W^\dagger}\biggr].
\end{equation}
Inserting $W^\dagger W=I$ under the trace gives, without approximation,
\begin{equation}\label{eq:tanh}
I_L(s,T)=(-1)^{\Sigma\cdot\gamma}\frac{Z_+-Z_-}{Z_++Z_-}
=(-1)^{\Sigma\cdot\gamma}\tanh\frac{F_\Sigma(s,T)}{2},
\end{equation}
where the dimensionless sector free-energy splitting is
\begin{equation}
F_\Sigma(s,T)\coloneqq\ln(Z_+/Z_-).
\end{equation}

Equation~(\ref{eq:tanh}) reduces quantization to the large-$L$ behavior of
the two sector partition functions. At low temperature on the
$s<\tfrac12$ side, a positive winding-line cost $\kappa$ gives
$Z_-/Z_+=O(\operatorname{poly}(L)e^{-\kappa L})$, hence
$F_\Sigma\to+\infty$. At high temperature, a positive domain-wall surface
tension $\sigma$ makes the sector asymmetry $O(e^{-\sigma L^2})$, hence
$F_\Sigma\to0$. Thus the index approaches the braiding sign in the SPT phase
and zero in the thermally disordered phase; exact reflection gives
$F_\Sigma\to-\infty$ and $\mathcal I=+1$ in the projected-paramagnetic phase.
At general finite temperature, the stabilizer-basis expansion in each
membrane sector $m=\pm1$ is a signed sum over closed $\mathbb Z_2$ spacetime
worldsheets of two cell-dual species. After isolating their exact mod-$2$
intersection sign, we denote the two resulting positive worldsheet ensembles
by copies $A$ and $B$;
here ``copy'' labels a worldsheet species, not a replica of the density
matrix. Copy $A$ carries $m$, and copy $B$ is its cell-dual partner. The
intersection sign differs from $+1$ only if at least one worldsheet has
nontrivial purely spatial homology. Let $Z_{A,m}^{(a_s)}$ and
$Z_B^{(b_s)}$ be the positive partition sums at fixed purely spatial homology
classes $a_s$ and $b_s$, and let
$Z_{A,m}\coloneqq\sum_{a_s}Z_{A,m}^{(a_s)}$ and
$Z_B\coloneqq\sum_{b_s}Z_B^{(b_s)}$. Define the total relative weights of
nontrivial spatial sheets by
\begin{equation}
\epsilon_{A,m}\coloneqq
\frac{\sum_{a_s\ne0}Z_{A,m}^{(a_s)}}{Z_{A,m}},\qquad
\epsilon_B\coloneqq
\frac{\sum_{b_s\ne0}Z_B^{(b_s)}}{Z_B}.
\end{equation}
Hypothesis (H) is the uniform area-law bound
\begin{equation}\label{eq:H-letter}
\epsilon_{A,m}+\epsilon_B\le Ce^{-\tau_{\rm sp}L^2},
\qquad m=\pm1 .
\end{equation}
For each fixed $(s,T)$ where the hypothesis is invoked,
$0<C<\infty$ and $\tau_{\rm sp}>0$ are independent of both $L$ and $m$. Thus Hypothesis
(H) is the physical assumption that the spatial-sheet sectors capable of
carrying the intersection sign have exponentially small relative weight in
both copies. Consequently,
$F_\Sigma=\ln(Z_{A,+}/Z_{A,-})+O(e^{-\tau_{\rm sp}L^2})$; positive
winding-line and interface tensions in copy $A$ then give the quantized phase
values. The derivation is in Sec.~\ref{supp:sm:interior} of the SM.

\textit{Numerical status.}---Assuming Hypothesis (H), one-copy QMC gives the
conditional $L=8$ map in Fig.~\ref{fig:maps}(b). Without Hypothesis (H),
direct two-copy QMC gives the five $L=4$ estimates in
Table~\ref{tab:hfree-letter}. Here $f$ labels the Trotter refinement defined
in Sec.~\ref{supp:sm:index-numerics} of the SM.

\begin{table}[t]
\caption{Direct QMC estimates of $I_{L=4}$.}
\label{tab:hfree-letter}
\begin{ruledtabular}
\begin{tabular}{c@{\quad}ccc}
$(s,T)$ & $f=1$ & $f=2$ & $f=4$ \\ \hline
$(0.30,0.30)$ & $-0.99999957$ & $-0.99999944$ & $-0.99999942$ \\
$(0.30,1.20)$ & $-0.00475$ & $\phantom{-}0.00298$ & --
\end{tabular}
\end{ruledtabular}
\end{table}

\textit{Full-cycle projection and Kennedy--Tasaki duality.}---The
$G_{\rm full}$-strong state $\rho_+$ fixes all six noncontractible plane
representatives. At most six logical $Z$ loops, of total length $6L$, map any
sector to all $+$ while changing at most $12L$ terms of $H(s)$. Since the local
couplings sum to one, the sector-Hamiltonian norm difference is at most $12L$,
and the $\beta$-Lipschitz bound gives
\begin{equation}\label{eq:KTbound}
0\le\ln Z_{\rm loc}(s,T)
-\ln\Tr[P_+e^{-\beta H(s)}]\le6\ln2+12\beta L .
\end{equation}
Thus the bulk free-energy densities agree up to $O(L^{-2})$. For the
non-invertible Kennedy--Tasaki map $\tilde D$ (the projected gauging map defined
in the SM)~\cite{KT,LOZ,CLY,Li:2024eih,KYH}, the fusion projector is $P_+$ and
$\tilde D^\dagger\tilde D=P_+$~\cite{Companion}; hence
$\Tr(e^{-\beta H(s)}P_+)=\Tr(e^{-\beta\tilde H(s)}P_+)$
for
$\tilde H(s)\coloneqq-(1-s)(\sum_lB_l+\sum_pB_p)
-s(\sum_lX_l+\sum_pX_p)$;
see Sec.~\ref{supp:sm:exact} of the SM. The projected KT pair
therefore has identical bulk free-energy singular loci, while $P_+$ fixes the
membrane index trivially to $+1$.

\textit{Discussion.}---In summary, the symmetry-projected three-dimensional
cluster interpolation supports thermal SPT, projected-paramagnetic, and
thermally disordered phases, with the predicted membrane-index values
$\mathcal I=-1,+1,0$, respectively. The
ensemble dependence suggests treating thermally
active global charge sectors as phase data in constrained mixed-state
classifications~\cite{LBC,LiYao}.
The direct two-copy reconstruction removes Hypothesis (H) for the reported
$L=4$ representative points only. Establishing phase-wide thermodynamic
quantization still requires multi-size and time-continuum control of the
sector asymptotics.

Decoded Wilson-loop correlations distinguish an unprojected toric-code Gibbs
transition even without exact higher-form symmetries and are
quasi-local-channel invariant~\cite{GPTC}. It remains open whether
$\langle S_\Sigma^{(1)}\rangle_{\rm tw}$ is quantized for general mixed states
with strong contractible one-form symmetry. Our projected-Gibbs derivation
converts it into a sector-partition-function ratio controlled by winding-line,
interface, and spatial-sheet tensions. Frameworks based on tensor networks,
channels, holography, or
axioms~\cite{Wang:2023uoj,Ellison:2024svg,Guo:2024iii,Sohal:2024qvq,Sun:2024iwm,Yang:2025pke,Luo:2025phx,Schafer-Nameki:2025fiy,Qi:2025tal}
need not share this structure, so strong symmetry alone may not quantize the
ratio. Extension requires an intrinsic flux twist stable under locally
reversible, symmetry-preserving channels~\cite{Sang:2023rsp,Sang:2025ssd}.

\textit{Acknowledgments.---}We thank Takamasa Ando, Ryohei Kobayashi, Linhao
Li, Ken Shiozaki, and Ryan Thorngren for helpful discussions.

\bibliography{refs}

@article{KT,
  author = {Kennedy, Tom and Tasaki, Hal},
  title = {Hidden {$\mathbb{Z}_2\times\mathbb{Z}_2$} symmetry breaking in {H}aldane-gap antiferromagnets},
  journal = {Phys. Rev. B},
  volume = {45},
  pages = {304},
  year = {1992},
  doi = {10.1103/PhysRevB.45.304}
}

@article{RBH,
  author = {Raussendorf, Robert and Bravyi, Sergey and Harrington, Jim},
  title = {Long-range quantum entanglement in noisy cluster states},
  journal = {Phys. Rev. A},
  volume = {71},
  pages = {062313},
  year = {2005},
  doi = {10.1103/PhysRevA.71.062313}
}

@article{LOZ,
  author = {Li, Linhao and Oshikawa, Masaki and Zheng, Yunqin},
  title = {Noninvertible duality transformation between symmetry-protected topological and spontaneously symmetry-broken phases},
  journal = {Phys. Rev. B},
  volume = {108},
  pages = {214429},
  year = {2023},
  eprint = {2301.07899},
  archivePrefix = {arXiv},
  doi = {10.1103/PhysRevB.108.214429}
}

@article{CLY,
  author = {Cao, Weiguang and Li, Linhao and Yamazaki, Masahito},
  title = {Generating lattice non-invertible symmetries},
  journal = {SciPost Phys.},
  volume = {17},
  pages = {104},
  year = {2024},
  eprint = {2406.05454},
  archivePrefix = {arXiv},
  doi = {10.21468/SciPostPhys.17.4.104}
}

@article{KYH,
  author = {Kim, Jintae and You, Yizhi and Han, Jung Hoon},
  title = {Noninvertible symmetry and topological holography for modulated {SPT} in one dimension},
  journal = {SciPost Phys.},
  volume = {19},
  pages = {110},
  year = {2025},
  eprint = {2507.02324},
  archivePrefix = {arXiv},
  doi = {10.21468/SciPostPhys.19.4.110}
}

@article{RYKB,
  author = {Roberts, Sam and Yoshida, Beni and Kubica, Aleksander and Bartlett, Stephen D.},
  title = {Symmetry-protected topological order at nonzero temperature},
  journal = {Phys. Rev. A},
  volume = {96},
  pages = {022306},
  year = {2017},
  eprint = {1611.05450},
  archivePrefix = {arXiv},
  doi = {10.1103/PhysRevA.96.022306}
}

@article{LBC,
  author = {Li, Linhao and Bi, Zhen and Cao, Weiguang},
  title = {Generalized symmetry-protected topological phases in mixed states from gauging dualities},
  journal = {arXiv preprint},
  eprint = {2603.17282},
  archivePrefix = {arXiv},
  year = {2026}
}

@article{PT,
  author = {Pollmann, Frank and Turner, Ari M.},
  title = {Detection of symmetry-protected topological phases in one dimension},
  journal = {Phys. Rev. B},
  volume = {86},
  pages = {125441},
  year = {2012},
  eprint = {1204.0704},
  archivePrefix = {arXiv},
  doi = {10.1103/PhysRevB.86.125441}
}

@article{GPTC,
  author = {Watanabe, Haruki},
  title = {Phase distinction of {G}ibbs states without symmetry breaking: topological invariants of the {3D} toric code},
  journal = {arXiv preprint},
  eprint = {2607.00134},
  archivePrefix = {arXiv},
  year = {2026}
}

@article{Wegner,
  author = {Wegner, Franz J.},
  title = {Duality in generalized {I}sing models and phase transitions without local order parameters},
  journal = {J. Math. Phys.},
  volume = {12},
  pages = {2259},
  year = {1971},
  doi = {10.1063/1.1665530}
}

@article{AizenmanBarskyFernandez,
  author = {Aizenman, Michael and Barsky, David J. and Fern{\'a}ndez, Roberto},
  title = {The phase transition in a general class of {Ising}-type models is sharp},
  journal = {J. Stat. Phys.},
  volume = {47},
  pages = {343--374},
  year = {1987},
  doi = {10.1007/BF01007515}
}

@article{LebowitzPfister,
  author = {Lebowitz, Joel L. and Pfister, Charles-Edouard},
  title = {Surface tension and phase coexistence},
  journal = {Phys. Rev. Lett.},
  volume = {46},
  pages = {1031--1033},
  year = {1981},
  doi = {10.1103/PhysRevLett.46.1031}
}

@article{ReissSchmidt,
  author = {Reiss, David A. and Schmidt, Kai Phillip},
  title = {Quantum robustness and phase transitions of the {3D} toric code in a field},
  journal = {SciPost Phys.},
  volume = {6},
  pages = {078},
  year = {2019},
  doi = {10.21468/SciPostPhys.6.6.078}
}

@misc{ParaToric,
  author = {Linsel, Simon M. and Pollet, Lode},
  title = {{ParaToric} 1.0: Continuous-time quantum {M}onte {C}arlo for the toric code in a parallel field},
  note = {arXiv:2510.14781},
  year = {2025}
}

@article{Li:2024eih,
    author = "Li, Linhao and Oshikawa, Masaki and Yan, Han",
    title = "{Generalized Kramers-Wanier Duality from Bilinear Phase Map}",
    eprint = "2403.16017",
    archivePrefix = "arXiv",
    primaryClass = "cond-mat.str-el",
    doi = "10.1103/g1l7-bvtf",
    journal = "Phys. Rev. Lett.",
    volume = "136",
    number = "24",
    pages = "240403",
    year = "2026"
}

@article{Ma:2022pvq,
    author = "Ma, Ruochen and Wang, Chong",
    title = "{Average Symmetry-Protected Topological Phases}",
    eprint = "2209.02723",
    archivePrefix = "arXiv",
    primaryClass = "cond-mat.str-el",
    doi = "10.1103/PhysRevX.13.031016",
    journal = "Phys. Rev. X",
    volume = "13",
    number = "3",
    pages = "031016",
    year = "2023"
}

@article{deGroot:2021vdi,
    author = "de Groot, Caroline and Turzillo, Alex and Schuch, Norbert",
    title = "{Symmetry Protected Topological Order in Open Quantum Systems}",
    eprint = "2112.04483",
    archivePrefix = "arXiv",
    primaryClass = "quant-ph",
    doi = "10.22331/q-2022-11-10-856",
    journal = "Quantum",
    volume = "6",
    pages = "856",
    year = "2022"
}

@misc{Companion,
  author = {Cao, Weiguang and Watanabe, Haruki},
  title = {Distinct finite-temperature phase diagrams of non-invertible {K}ennedy--{T}asaki duals},
  eprint = {2607.24231},
  archivePrefix = {arXiv},
  primaryClass = {cond-mat.str-el},
  year = {2026}
}

@article{LiYao,
  author = {Li, Linhao and Yao, Yuan},
  title = {Mixed-state topological phase: Quantized topological order parameter and {Lieb--Schultz--Mattis} theorem},
  journal = {arXiv preprint},
  year = {2026},
  eprint = {2603.24031},
  archivePrefix = {arXiv}
}

@article{FerrenbergXuLandau,
  author = {Ferrenberg, A. M. and Xu, J. and Landau, D. P.},
  title = {Pushing the limits of Monte Carlo simulations for the three-dimensional Ising model},
  journal = {Phys. Rev. E},
  volume = {97},
  pages = {043301},
  year = {2018},
  doi = {10.1103/PhysRevE.97.043301},
  eprint = {1806.03558},
  archivePrefix = {arXiv}
}

@article{KPSDV,
  author = {Kos, F. and Poland, D. and Simmons-Duffin, D. and Vichi, A.},
  title = {Precision islands in the Ising and $O(N)$ models},
  journal = {JHEP},
  volume = {2016},
  number = {08},
  pages = {036},
  year = {2016},
  doi = {10.1007/JHEP08(2016)036},
  eprint = {1603.04436},
  archivePrefix = {arXiv}
}

@article{PhysRevB.80.155131,
  title = {Tensor-entanglement-filtering renormalization approach and symmetry-protected topological order},
  author = {Gu, Zheng-Cheng and Wen, Xiao-Gang},
  journal = {Phys. Rev. B},
  volume = {80},
  issue = {15},
  pages = {155131},
  numpages = {23},
  year = {2009},
  month = {Oct},
  publisher = {American Physical Society},
  doi = {10.1103/PhysRevB.80.155131},
  url = {https://link.aps.org/doi/10.1103/PhysRevB.80.155131}
}

@article{Pollmann:2009ryx,
    author = "Pollmann, Frank and Turner, Ari M. and Berg, Erez and Oshikawa, Masaki",
    title = "{Entanglement spectrum of a topological phase in one dimension}",
    eprint = "0910.1811",
    archivePrefix = "arXiv",
    primaryClass = "cond-mat.str-el",
    doi = "10.1103/PhysRevB.81.064439",
    journal = "Phys. Rev. B",
    volume = "81",
    number = "6",
    pages = "064439",
    year = "2010"
}

@article{Pollmann:2009mhk,
    author = "Pollmann, Frank and Berg, Erez and Turner, Ari M. and Oshikawa, Masaki",
    title = "{Symmetry protection of topological phases in one-dimensional quantum spin systems}",
    eprint = "0909.4059",
    archivePrefix = "arXiv",
    primaryClass = "cond-mat.str-el",
    doi = "10.1103/PhysRevB.85.075125",
    journal = "Phys. Rev. B",
    volume = "85",
    number = "7",
    pages = "075125",
    year = "2012"
}

@article{Chen:2010zpc,
    author = "Chen, Xie and Gu, Zheng-Cheng and Wen, Xiao-Gang",
    title = "{Classification of gapped symmetric phases in one-dimensional spin systems}",
    eprint = "1008.3745",
    archivePrefix = "arXiv",
    primaryClass = "cond-mat.str-el",
    doi = "10.1103/PhysRevB.83.035107",
    journal = "Phys. Rev. B",
    volume = "83",
    number = "3",
    pages = "035107",
    year = "2011"
}

@article{PhysRevB.84.235128,
  title = {Complete classification of one-dimensional gapped quantum phases in interacting spin systems},
  author = {Chen, Xie and Gu, Zheng-Cheng and Wen, Xiao-Gang},
  journal = {Phys. Rev. B},
  volume = {84},
  issue = {23},
  pages = {235128},
  numpages = {14},
  year = {2011},
  month = {Dec},
  publisher = {American Physical Society},
  doi = {10.1103/PhysRevB.84.235128},
  url = {https://link.aps.org/doi/10.1103/PhysRevB.84.235128}
}

@article{Chen:2011pg,
    author = "Chen, Xie and Gu, Zheng-Cheng and Liu, Zheng-Xin and Wen, Xiao-Gang",
    title = "{Symmetry protected topological orders and the group cohomology of their symmetry group}",
    eprint = "1106.4772",
    archivePrefix = "arXiv",
    primaryClass = "cond-mat.str-el",
    doi = "10.1103/PhysRevB.87.155114",
    journal = "Phys. Rev. B",
    volume = "87",
    number = "15",
    pages = "155114",
    year = "2013"
}

@article{Lessa:2024gcw,
    author = "Lessa, Leonardo A. and Ma, Ruochen and Zhang, Jian-Hao and Bi, Zhen and Cheng, Meng and Wang, Chong",
    title = "{Strong-to-Weak Spontaneous Symmetry Breaking in Mixed Quantum States}",
    eprint = "2405.03639",
    archivePrefix = "arXiv",
    primaryClass = "quant-ph",
    doi = "10.1103/PRXQuantum.6.010344",
    journal = "PRX Quantum",
    volume = "6",
    number = "1",
    pages = "010344",
    year = "2025"
}

@article{Castelnovo:2008jyy,
    author = "Castelnovo, Claudio and Chamon, Claudio",
    title = "{Topological order in a three-dimensional toric code at finite temperature}",
    eprint = "0804.3591",
    archivePrefix = "arXiv",
    primaryClass = "cond-mat.str-el",
    doi = "10.1103/PhysRevB.78.155120",
    journal = "Phys. Rev. B",
    volume = "78",
    number = "15",
    pages = "155120",
    year = "2008"
}

@misc{Negari:2025jdc,
    author = "Negari, Amir-Reza and Lessa, Leonardo A. and Sahu, Subhayan",
    title = "{Symmetry enforces entanglement at high temperatures}",
    eprint = "2508.20166",
    archivePrefix = "arXiv",
    primaryClass = "quant-ph",
    month = "8",
    year = "2025"
}

@article{Hastings:2011iig,
    author = "Hastings, Matthew B.",
    title = "{Topological Order at Nonzero Temperature}",
    eprint = "1106.6026",
    archivePrefix = "arXiv",
    primaryClass = "quant-ph",
    doi = "10.1103/PhysRevLett.107.210501",
    journal = "Phys. Rev. Lett.",
    volume = "107",
    number = "21",
    pages = "210501",
    year = "2011"
}

@article{Sang:2023rsp,
    author = "Sang, Shengqi and Zou, Yijian and Hsieh, Timothy H.",
    title = "{Mixed-State Quantum Phases: Renormalization and Quantum Error Correction}",
    eprint = "2310.08639",
    archivePrefix = "arXiv",
    primaryClass = "quant-ph",
    doi = "10.1103/PhysRevX.14.031044",
    journal = "Phys. Rev. X",
    volume = "14",
    number = "3",
    pages = "031044",
    year = "2024"
}

@article{Ellison:2024svg,
    author = "Ellison, Tyler D. and Cheng, Meng",
    title = "{Toward a Classification of Mixed-State Topological Orders in Two Dimensions}",
    eprint = "2405.02390",
    archivePrefix = "arXiv",
    primaryClass = "cond-mat.str-el",
    doi = "10.1103/PRXQuantum.6.010315",
    journal = "PRX Quantum",
    volume = "6",
    number = "1",
    pages = "010315",
    year = "2025"
}

@article{Guo:2024iii,
    author = "Guo, Yuchen and Zhang, Jian-Hao and Zhang, Hao-Ran and Yang, Shuo and Bi, Zhen",
    title = "{Locally Purified Density Operators for Symmetry-Protected Topological Phases in Mixed States}",
    eprint = "2403.16978",
    archivePrefix = "arXiv",
    primaryClass = "cond-mat.str-el",
    doi = "10.1103/PhysRevX.15.021060",
    journal = "Phys. Rev. X",
    volume = "15",
    number = "2",
    pages = "021060",
    year = "2025"
}

@article{Sohal:2024qvq,
    author = "Sohal, Ramanjit and Prem, Abhinav",
    title = "{Noisy Approach to Intrinsically Mixed-State Topological Order}",
    eprint = "2403.13879",
    archivePrefix = "arXiv",
    primaryClass = "cond-mat.str-el",
    doi = "10.1103/PRXQuantum.6.010313",
    journal = "PRX Quantum",
    volume = "6",
    number = "1",
    pages = "010313",
    year = "2025"
}

@article{Wang:2023uoj,
    author = "Wang, Zijian and Wu, Zhengzhi and Wang, Zhong",
    title = "{Intrinsic Mixed-State Topological Order}",
    eprint = "2307.13758",
    archivePrefix = "arXiv",
    primaryClass = "quant-ph",
    doi = "10.1103/PRXQuantum.6.010314",
    journal = "PRX Quantum",
    volume = "6",
    number = "1",
    pages = "010314",
    year = "2025"
}

@article{Sun:2024iwm,
    author = "Sun, Shijun and Zhang, Jian-Hao and Bi, Zhen and You, Yizhi",
    title = "{Holographic View of Mixed-State Symmetry-Protected Topological Phases in Open Quantum Systems}",
    eprint = "2410.08205",
    archivePrefix = "arXiv",
    primaryClass = "quant-ph",
    doi = "10.1103/PRXQuantum.6.020333",
    journal = "PRX Quantum",
    volume = "6",
    number = "2",
    pages = "020333",
    year = "2025"
}

@misc{Sang:2025ssd,
    author = "Sang, Shengqi and Lessa, Leonardo A. and Mong, Roger S. K. and Grover, Tarun and Wang, Chong and Hsieh, Timothy H.",
    title = "{Mixed-state phases from local reversibility}",
    eprint = "2507.02292",
    archivePrefix = "arXiv",
    primaryClass = "quant-ph",
    month = "7",
    year = "2025"
}

@misc{Yang:2025pke,
    author = "Yang, Tai-Hsuan and Shi, Bowen and Lee, Jong Yeon",
    title = "{Topological Mixed States: Phases of Matter from Axiomatic Approaches}",
    eprint = "2506.04221",
    archivePrefix = "arXiv",
    primaryClass = "cond-mat.str-el",
    month = "6",
    year = "2025"
}

@article{Luo:2025phx,
    author = "Luo, Ran and Wang, Yi-Nan and Bi, Zhen",
    title = "{Topological Holography for Mixed-State Phases and Phase Transitions}",
    eprint = "2507.06218",
    archivePrefix = "arXiv",
    primaryClass = "cond-mat.str-el",
    doi = "10.1103/9kmh-gjf8",
    journal = "PRX Quantum",
    volume = "6",
    number = "4",
    pages = "040358",
    year = "2025"
}

@misc{Schafer-Nameki:2025fiy,
    author = "Schafer-Nameki, Sakura and Tiwari, Apoorv and Warman, Alison and Zhang, Carolyn",
    title = "{SymTFT Approach for Mixed States with Non-Invertible Symmetries}",
    eprint = "2507.05350",
    archivePrefix = "arXiv",
    primaryClass = "quant-ph",
    month = "7",
    year = "2025"
}

@misc{Qi:2025tal,
    author = "Qi, Marvin and Sohal, Ramanjit and Chen, Xie and Stephen, David T. and Prem, Abhinav",
    title = "{The Symmetry Taco: Equivalences between Gapped, Gapless, and Mixed-State SPTs}",
    eprint = "2507.05335",
    archivePrefix = "arXiv",
    primaryClass = "cond-mat.str-el",
    month = "7",
    year = "2025"
}

\clearpage
\onecolumngrid
\begin{center}
{\large\bfseries Supplemental Material}\\[4pt]
{\large\bfseries Quantized topological invariant of symmetry-projected Gibbs states}\\[8pt]
Weiguang Cao and Haruki Watanabe
\end{center}
\vspace{1.2em}
\twocolumngrid

\setcounter{secnumdepth}{3}
\setcounter{section}{0}
\setcounter{equation}{0}
\setcounter{figure}{0}
\setcounter{table}{0}
\renewcommand{\thesection}{S\arabic{section}}
\renewcommand{\theequation}{S\arabic{equation}}
\renewcommand{\thefigure}{S\arabic{figure}}
\renewcommand{\thetable}{S\arabic{table}}
\renewcommand{\theHsection}{S\arabic{section}}
\renewcommand{\theHequation}{S\arabic{equation}}
\renewcommand{\theHfigure}{S\arabic{figure}}
\renewcommand{\theHtable}{S\arabic{table}}

This Supplemental Material follows the order of the Letter.  First,
Sec.~\ref{supp:sm:setup} fixes the ensembles and notation, and
Secs.~\ref{supp:sm:fusion}--\ref{supp:sm:s1-loops} derive the fusion rule and endpoint
mappings.  Sections~\ref{supp:sm:proj-qmc} and \ref{supp:sm:tstar-scaling} then describe
the finite-regulator phase diagram.  Section~\ref{supp:sm:s1-twist} defines the
twisted membrane and proves the exact boundary-line results, after which
Sec.~\ref{supp:sm:interior} states the spatial-sheet-suppression assumption
[Hypothesis (H)] and derives conditional interior quantization.  Finally,
Sec.~\ref{supp:sm:stw-num} gives the one-copy proxy and sheet-cost calculation,
and Sec.~\ref{supp:sm:hfree-qmc} gives the separate direct finite-$L$ calculation
that retains all sheet classes.

\section{Setup: models, symmetries, and projected ensembles}
\label{supp:sm:setup}

We set $k_B=1$, write $\beta\coloneqq1/T$, and measure energy in the unit
coupling of the endpoint Hamiltonians.  The interpolation parameter obeys
$0\le s\le1$.  Otherwise we use the conventions of the Letter and the
companion Letter~\cite{Companion}. The Raussendorf--Bravyi--Harrington (RBH) model is
defined on a cubic $3$-torus of linear size $L$, with qubits on the
$N_l\coloneqq3L^3$ links and $3L^3$ plaquettes. The dressings are
$B_l\coloneqq\prod_{p\ni l}Z_p$ and
$B_p\coloneqq\prod_{l\in\partial p}Z_l$, the cluster stabilizers are
$K_p\coloneqq X_pB_p$ and $K_l\coloneqq X_lB_l$, and the contractible one-form
generators are $S^{(1)}_c\coloneqq\prod_{p\in\partial c}X_p$ and
$S^{(1)}_v\coloneqq\prod_{l\ni v}X_l$.  The two Hamiltonian interpolations
used below are
\begin{equation}
\begin{aligned}
H(s)&\coloneqq-(1-s)\!\left(\sum_lK_l+\sum_pK_p\right)\\[-2pt]
&\hspace{2.2em}-s\!\left(\sum_lX_l+\sum_pX_p\right),\\
\tilde H(s)&\coloneqq-(1-s)\!\left(\sum_lB_l+\sum_pB_p\right)\\[-2pt]
&\hspace{2.2em}-s\!\left(\sum_lX_l+\sum_pX_p\right).
\end{aligned}
\label{supp:eq:sm-hamiltonians}
\end{equation}
Let $\langle l,p\rangle_\partial$ denote an incident link--plaquette pair,
$l\in\partial p$. The finite-depth cluster entangler is
\begin{equation}
U\coloneqq\prod_{\langle l,p\rangle_\partial}CZ_{l,p},
\end{equation}
where $CZ_{l,p}$ is a controlled-$Z$ gate. The gauging map $D$ is
defined in Sec.~\ref{supp:sm:fusion}, and
$\tilde D\coloneqq UDU^\dagger$ is the Kennedy--Tasaki map.  For a
noncontractible spatial two-cycle $\Sigma$ and a noncontractible one-cycle
$\gamma$, we write
$S_\Sigma^{(1)}\coloneqq\prod_{p\in\Sigma}X_p$ and denote their mod-$2$
intersection number by
$\Sigma\mathbin{\cdot}\gamma\coloneqq
\langle[\Sigma],[\gamma]\rangle\in\mathbb Z_2$.  The projectors
$P_{\rm loc}$ and $P_+$ are defined next.

For clarity, we distinguish the two projected ensembles explicitly. Let
\begin{equation}
\begin{aligned}
G_{\rm loc}&\coloneqq\big\langle\{S^{(1)}_c\}_c,\{S^{(1)}_v\}_v\big\rangle ,\\
P_{\rm loc}&\coloneqq\frac{1}{|G_{\rm loc}|}\sum_{g\in G_{\rm loc}}g,\\
G_{\rm full}&\coloneqq\big\langle G_{\rm loc},S^{\rm nc}_1,\ldots,S^{\rm nc}_6\big\rangle ,\\
P_+&\coloneqq\frac{1}{|G_{\rm full}|}\sum_{g\in G_{\rm full}}g=P_{\rm loc}\prod_{a=1}^{6}\frac{1+S_a^{\rm nc}}{2}.
\end{aligned}
\label{supp:eq:ensemble-glossary}
\end{equation}
Thus $G_{\rm loc}$ contains the contractible cube and vertex generators, whereas $G_{\rm full}/G_{\rm loc}\simeq\mathbb Z_2^6$ records the six noncontractible generators. Products are understood over $\mathbb Z_2$: two parallel representatives of the same noncontractible plane class combine to a contractible element of $G_{\rm loc}$. With
$Z_{\rm loc}(s)\coloneqq\Tr[P_{\rm loc}e^{-\beta H(s)}]$ and
$Z_+(s)\coloneqq\Tr[P_+e^{-\beta H(s)}]$, define
\begin{equation}
\rho_{\rm loc}(s)\coloneqq
\frac{P_{\rm loc}e^{-\beta H(s)}P_{\rm loc}}{Z_{\rm loc}(s)},\qquad
\rho_+(s)\coloneqq\frac{P_+e^{-\beta H(s)}P_+}{Z_+(s)}.
\label{supp:eq:rho-glossary}
\end{equation}
At the RBH fixed point, $\rho_{\rm loc}(0)$ is exactly the hard-projection
ensemble of Ref.~\cite{RYKB}; along the interpolation we use the same
contractible-generator projection and denote it by $\rho_{\rm loc}(s)$.  It
sums over the six noncontractible charges.  By contrast, $\rho_+(s)$ fixes all
six of them to $+1$ and is the all-$+$ full-fusion state.  Throughout this
Supplemental Material, ``strongly symmetric'' without another qualifier
means strong with respect to $G_{\rm loc}$ and refers to $\rho_{\rm loc}$;
$P_+$ or a fixed noncontractible sector is always named explicitly.

Finite-range locality relates the two symmetry groups at the Hamiltonian
level.  For any local Pauli term and any noncontractible plane
$S_a^{\rm nc}$, choose a disjoint parallel representative
$\widehat S_a^{\rm nc}$.  Their product is contractible, whereas
$\widehat S_a^{\rm nc}$ acts trivially on the local term.  Contractible
invariance therefore implies $[H,S_a^{\rm nc}]=0$ once $L$ exceeds the
interaction range.  Thus $\rho_{\rm loc}$ is weakly $G_{\rm full}$ symmetric,
$S_a^{\rm nc}\rho_{\rm loc}(S_a^{\rm nc})^\dagger=\rho_{\rm loc}$, but is not
strongly $G_{\rm full}$ symmetric because $P_{\rm loc}$ sums rather than fixes
the noncontractible charges.

\section{Exact projectors and endpoint mappings}
\label{supp:sm:exact}

\subsection{The fusion rule
\texorpdfstring{$\tilde D^\dagger\tilde D=P_+$}{D-tilde dagger D-tilde = P+},
and \texorpdfstring{$P_+$ versus $P_{\rm loc}$}{P+ versus P-loc}}
\label{supp:sm:fusion}
The gauging map of the dressings admits the bilinear phase-map representation~\cite{Li:2024eih}
\begin{equation}
D=\mathcal N^{-1}\!\!\sum_{\{\sigma\},\{\sigma'\}}\!(-1)^{\sum_j a_j(\sigma)\,\sigma'_j}\,|\{\sigma\}\rangle\langle\{\sigma'\}|,
\end{equation}
with $a_j(\sigma)\coloneqq(M\sigma)_j$, where $|\{\sigma\}\rangle$ is the
computational ($Z$) basis on the $N\coloneqq6L^3$ qubits,
$M_{ij}\coloneqq1$ iff $Z_j$ appears in the dressing $B_i$ and $0$ otherwise
($M=M^\top$, the link--plaquette incidence matrix of the $3$-torus), and
$\mathcal N\coloneqq(2^{N}|\ker M|)^{1/2}$. Contracting the intermediate
computational-basis state gives
\begin{align}
(D^\dagger D)_{\{\sigma'\}\{\sigma''\}}
&=\mathcal N^{-2}\!\sum_{\{\sigma\}}(-1)^{\sigma\cdot M(\sigma'+\sigma'')}\nonumber\\
&=\frac{2^{N}}{\mathcal N^{2}}\,\delta_{\sigma'+\sigma''\in\ker M},
\end{align}
i.e.\ $D^\dagger D=|\ker M|^{-1}\sum_{\bm n\in\ker M}X^{\bm n}$
[$X^{\bm n}\coloneqq\prod_jX_j^{n_j}$ for $\bm n\in\{0,1\}^N$, similarly
$Z^{\bm m}$; $M$ acts over $\mathbb F_2$], the group average over
$\{X^{\bm n}\}_{\bm n\in\ker M}$---a projector, because the set is a group
(the \emph{fusion group}). The entangler fixes every element of this group
(shown below); hence
$\tilde D^\dagger\tilde D=UD^\dagger DU^\dagger=D^\dagger D=P_+$.
What makes $P_+$ the \emph{full} fusion projector is the content of $\ker M$. The condition $M\bm n=0$ splits by species. The plaquette part of $\bm n$ must overlap every $B_l$ evenly---the closed ($2$-cycle) plaquette sets, generated by the contractible cube generators ($2^{L^3-1}$ elements) together with the three noncontractible plane classes---while the link part must overlap every $B_p$ evenly, generated by the contractible vertex stars ($2^{L^3-1}$) together with the three noncontractible dual planes. Hence
\begin{equation}\label{supp:eq:Pplus-Ploc}
P_+=P_{\rm loc}\prod_{a=1}^{6}\tfrac12\big(1+S_a^{\rm nc}\big),
\end{equation}
with $S_a^{\rm nc}$ representatives of the six noncontractible plane generators; the fusion rule enforces the noncontractible generators as well, an index-$2^6$ refinement of $P_{\rm loc}$. Thus $P_+$ fixes the six noncontractible generators in addition to the contractible ones: $S_\Sigma^{(1)}P_+=P_+$, whereas the membrane remains a thermal observable in $\rho_{\rm loc}$.

The entangler has a particularly simple action on these strings: it fixes an
$X$-string precisely when the string belongs to the fusion group. Conjugation
by the $CZ$ circuit gives $UX^{\bm n}U^\dagger=\pm X^{\bm n}Z^{M\bm n}$, and
both the dressing and the reordering sign are trivial exactly when
$M\bm n=0$. It follows that $U$ commutes with $P_{\rm loc}$, with $P_+$, and
with each noncontractible sector projector separately. Together with
$U^\dagger H(s)U=H(1-s)$, cyclicity gives the exact mirror relation
$Z_{\rm loc}(s,T)=Z_{\rm loc}(1-s,T)$, equally for $P_+$ and sector by sector
at finite $L$. Any coexistence line between the two low-temperature phases is
therefore pinned to $s=\tfrac12$.

We next compare the two ensembles without assuming which sector dominates.
They differ by only a subextensive free energy. Decomposing the local
projector over the $2^6$ noncontractible one-form charges,
\begin{equation}\label{supp:eq:Ploc-sectors}
P_{\rm loc}=\!\!\sum_{\varepsilon\in\mathbb F_2^6}\!\!P_\varepsilon,\qquad
P_\varepsilon\coloneqq P_{\rm loc}\prod_{a=1}^{6}
\tfrac12\big(1+(-1)^{\varepsilon_a}S_a^{\rm nc}\big),
\end{equation}
with $P_{\varepsilon=0}=P_+$ [Eq.~(\ref{supp:eq:Pplus-Ploc})], gives
$Z_{\rm loc}=\sum_{\varepsilon}Z_\varepsilon$, where
$Z_\varepsilon\coloneqq\Tr[P_\varepsilon e^{-\beta H(s)}]$. Each
$Z_\varepsilon\ge0$ is the trace of a projected Gibbs operator and $Z_+$ is
the $\varepsilon=0$ term; consequently, the lower bound
$Z_{\rm loc}\ge Z_+$ is immediate. The upper bound we obtain \emph{without}
assuming that the all-$+$ noncontractible sector dominates---which we have
not proved for the interacting interpolation---by transporting each sector
to $P_+$ with a unitary of bounded local cost. For each $\varepsilon$ pick at
most six noncontractible logical $Z$-loops, one per charge to be reversed;
their product $V_\varepsilon$ is unitary, supported on at most $6L$ qubits
(one length-$L$ loop per flipped charge), commutes with $P_{\rm loc}$ and
with the unflipped $S_b^{\rm nc}$, and anticommutes with each targeted
$S_a^{\rm nc}$, which implies
$V_\varepsilon P_+V_\varepsilon^\dagger=P_\varepsilon$. On each supported
link or plaquette qubit, $Z$ anticommutes with the corresponding $X$ and $XB$
terms. These are the only terms of
\[
\begin{aligned}
H(s)\coloneqq{}&-(1-s)\Big(\sum_lX_lB_l+\sum_pX_pB_p\Big)\\[-2pt]
&-s\Big(\sum_lX_l+\sum_pX_p\Big)
\end{aligned}
\]
that $V_\varepsilon$ fails to commute with. Since the two local couplings sum to $1$, each supported qubit contributes at most $2(1-s)+2s=2$ to the norm, yielding
\begin{equation}\label{supp:eq:Veps-norm}
\big\|V_\varepsilon H(s)V_\varepsilon^\dagger-H(s)\big\|\le 2\cdot 6L=12L .
\end{equation}
The log partition function is then $\beta$-Lipschitz in the Hamiltonian in
operator norm. The step rests on $P_+$ commuting with the whole interpolant:
$P_+$ commutes with $H(s)$, as does
$P_\varepsilon=V_\varepsilon P_+V_\varepsilon^\dagger$ (it is $P_{\rm loc}$
times a product of the $S_a^{\rm nc}$, each of which commutes with $H(s)$),
whence $[P_+,V_\varepsilon^\dagger H(s)V_\varepsilon]=0$ and therefore
$[P_+,H_t]=0$ along the segment
$H_t\coloneqq H(s)+t\Delta$, with
$\Delta\coloneqq V_\varepsilon^\dagger H(s)V_\varepsilon-H(s)$. Because $P_+$
commutes with $H_t$, the Duhamel integrand is independent of its
imaginary-time split parameter $r\in[0,1]$:
$e^{-r\beta H_t}P_+e^{-(1-r)\beta H_t}=P_+e^{-\beta H_t}$. For
$g(t)\coloneqq\ln\Tr[P_+e^{-\beta H_t}]$, this gives
$g'(t)=-\beta\,\Tr[R_t\Delta]$ with
$R_t\coloneqq P_+e^{-\beta H_t}\big/\Tr[P_+e^{-\beta H_t}]$ a bona fide
projected Gibbs state---$R_t\succeq0$ and $\Tr R_t=1$ manifestly, $P_+$ and
$e^{-\beta H_t}$ being commuting positive-semidefinite operators---whence
$|g'(t)|\le\beta\|\Delta\|$. Applying this along $t\in[0,1]$ and using
$Z_\varepsilon=\Tr[P_+e^{-\beta
V_\varepsilon^\dagger H(s)V_\varepsilon}]$ (cyclicity, with
$V_\varepsilon P_+V_\varepsilon^\dagger=P_\varepsilon$),
Eq.~(\ref{supp:eq:Veps-norm}) gives
$e^{-12\beta L}Z_+\le Z_\varepsilon\le e^{12\beta L}Z_+$. Summing the
$2^6$ sectors,
\begin{equation}\label{supp:eq:sector-bound}
0\;\le\;\ln Z_{\rm loc}-\ln Z_+\;\le\;6\ln2+12\beta L .
\end{equation}
unconditionally at every $s$, $\beta$, and $L$. Writing
$f_\alpha\coloneqq-(T/N)\ln Z_\alpha$ for
$\alpha\in\{\mathrm{loc},+\}$, with $N=6L^3$, this is the
free-energy-density gap
\begin{equation}\label{supp:eq:fdens-bound}
0\;\le\;f_+-f_{\rm loc}\;\le\;\frac{6T\ln2+12L}{6L^3}=O(L^{-2}).
\end{equation}
Whenever the thermodynamic free-energy-density limits exist,
Eq.~(\ref{supp:eq:fdens-bound}) makes the two limiting functions identical.
Therefore, the bulk free-energy densities of $\rho_{\rm loc}$ and $\rho_+$
and their singular loci coincide. The bound does not equate sector-sensitive observables,
strong-symmetry phase labels, or individual finite-$L$ data. In particular,
$S_\Sigma^{(1)}P_+=P_+$ pins the membrane in $\rho_+$, while it fluctuates
thermally in $\rho_{\rm loc}$ (Sec.~\ref{supp:sm:s1-twist}).

The projected identity quoted in the Letter follows directly from the same bilinear representation. Flipping $\sigma'_j$ (right multiplication by $X_j$) multiplies the phase by $(-1)^{(M\sigma)_j}$, the $B_j$ eigenvalue, while the $B_j$ eigenvalue of $\sigma'$ passes to a flip of $\sigma_j$; hence $DX_j=B_jD$ and $DB_j=X_jD$. Conjugating by the entangler, $\tilde D(X_jB_j)=B_j\tilde D$ and $\tilde DX_j=X_j\tilde D$, which is the intertwining relation $\tilde DH(s)=\tilde H(s)\tilde D$~\cite{Companion}. The bilinear representation also makes $D$ real symmetric; hence $DD^\dagger=D^\dagger D=P_+$ and likewise $\tilde D\tilde D^\dagger=\tilde D^\dagger\tilde D=P_+$. Combining the two with cyclicity,
\begin{equation}\label{supp:eq:proj-identity}
\Tr\big(P_+e^{-\beta H(s)}\big)=\Tr\big(\tilde De^{-\beta H(s)}\tilde D^\dagger\big)=\Tr\big(P_+e^{-\beta\tilde H(s)}\big).
\end{equation}
Equation~(\ref{supp:eq:proj-identity}) is the exact finite-$L$ KT identity for the all-$+$ full-fusion state $\rho_+$. The same noncontractible-sector decomposition applies to the dual family $\tilde H(s)$, whose two species decouple exactly at every $(s,T)$ and carry the positive worldsheet representation of Sec.~\ref{supp:sm:interior} with no dropped topological terms. The gauge-side simulations of Sec.~\ref{supp:sm:proj-qmc} instead use a finite-$\mu$ regulator approaching $\rho_{\rm loc}$. After the local-projection limit, Eq.~(\ref{supp:eq:sector-bound}) transfers the thermodynamic phase-transition locus, but no finite-$L$ observable, between $\rho_{\rm loc}$ and $\rho_+$.

\subsection{Exact finite-temperature transition of the locally projected
paramagnet at \texorpdfstring{$s=1$}{s=1}}
\label{supp:sm:s1-ising}
In the locally projected ensemble $\rho_{\rm loc}$ the ``trivial'' endpoint $s=1$ carries a genuine finite-temperature phase transition at the same temperature $T_c^{\rm proj}(0)$ that the mirror identity assigns to the cluster endpoint $s=0$. Before projection, its link species $H_1\coloneqq-\sum_l X_l$ (the plaquette species is an identical mirror copy) is a free paramagnet with $\mathrm{Tr}\,e^{-\beta H_1}=(2\cosh\beta)^{N_l}$, analytic at every temperature. The contractible-generator projector nevertheless maps each species exactly to a three-dimensional Ising model, as the following derivation shows.

For the derivation, consider a cubic lattice with
$N_v\coloneqq L^3$ vertices and $N_l=3L^3$ links carrying one qubit
each on the periodic $3$-torus. The star (Gauss) operator is
$S^{(1)}_v\coloneqq\prod_{l\ni v}X_l$, and
$P_{\rm vertex}\coloneqq\prod_v\tfrac{1}{2}(1+S^{(1)}_v)$ (the vertex-star
factor of $P_{\rm loc}$, as named in the Letter) projects onto the
contractible charge-free sector $S^{(1)}_v=+1$ for all $v$ while leaving the
noncontractible charges summed. The object of interest is
$Z_P(\beta)\coloneqq\mathrm{Tr}\big[e^{\beta\sum_l X_l}P_{\rm vertex}\big]$.
The cube factor $P_{\rm cube}$ of $P_{\rm loc}$ acts identically on the
decoupled plaquette species: plaquettes are the links, and cubes the vertices,
of the dual cubic lattice, making the cube operators the dual vertex stars.
The derivation below therefore maps the two-species locally projected trace
at $s=1$ to the product of two identical Ising partition functions with the
same singular temperature.

First, the projected trace is classical. Every operator in $Z_P$ is a
product of link $X$'s and therefore commutes with all the others. The trace
can be evaluated in the joint eigenbasis
$X_l|x\rangle=x_l|x\rangle$, $x_l=\pm1$. There $P_{\rm vertex}$ enforces the
local constraints $\prod_{l\ni v}x_l=+1$; the links with $x_l=-1$ must meet
every vertex an even number of times and hence form closed loops. Thus
$Z_P=e^{\beta N_l}\sum_{\mathcal C\,\mathrm{closed}}
e^{-2\beta|\mathcal C|}$ is a classical loop gas.

An alternative expansion makes the Ising representation explicit. Expand
the projector over the Gauss group,
\begin{equation}
P_{\rm vertex}=2^{-N_v}\sum_{S\subseteq\mathcal V}\prod_{v\in S}S^{(1)}_v ,
\end{equation}
with $\mathcal V$ the vertex set. The global relation
$\prod_vS_v^{(1)}=I$ makes $S$ and $\mathcal V\setminus S$ represent the
same Gauss-group element. Thus every distinct element occurs twice in this
subset sum, and $2^{-N_v}$ gives the normalized average over the
$2^{N_v-1}$ distinct elements. Since $X^2=1$, each term factorizes over
links, $\prod_{v\in S}S^{(1)}_v=\prod_l X_l^{\,n_l(S)}$, where
$n_l(S)\in\{0,1\}$ is the parity of the number of endpoints of $l$ contained
in $S$.

The resulting trace factorizes over links. Using
$\mathrm{tr}\,e^{\beta X}=2\cosh\beta$ and
$\mathrm{tr}\,[e^{\beta X}X]=2\sinh\beta$ per qubit,
\begin{equation}
\mathrm{Tr}\big[e^{\beta\sum_l X_l}\textstyle\prod_{v\in S}S^{(1)}_v\big]=(2\cosh\beta)^{N_l}\,(\tanh\beta)^{B(S)},
\end{equation}
with $B(S)$ the number of links having exactly one endpoint in $S$.

Now encode $S$ in classical vertex spins, $\sigma_v\coloneqq-1$ for
$v\in S$ and $+1$ otherwise, which gives a bijection between subsets and
spin configurations. Then
$B(S)=\sum_{\langle vv'\rangle}\tfrac12(1-\sigma_v\sigma_{v'})$ is the
domain-wall area, and
\begin{equation}\label{supp:eq:s1-ising}
Z_P(\beta)=c_\beta\sum_{\{\sigma\}}e^{K\sum_{\langle vv'\rangle}\sigma_v\sigma_{v'}},\qquad e^{-2K}\coloneqq\tanh\beta,
\end{equation}
with $c_\beta\coloneqq2^{-N_v}(2\cosh\beta)^{N_l}(\tanh\beta)^{N_l/2}$---the ferromagnetic nearest-neighbor Ising model on the simple cubic lattice of vertices ($S\to\mathcal V\setminus S$ is its global spin flip).

This exact representation also locates and characterizes the singularity.
The prefactor $c_\beta$ is analytic for $0<\beta<\infty$, and
$K=-\tfrac12\ln\tanh\beta$ is analytic in $\beta$ with
$dK/d\beta=-1/\sinh2\beta\neq0$. Hence the free-energy density of the
projected paramagnet has exactly the singularities of the three-dimensional
Ising model, located at
$K=K_c^{\rm Ising}\approx0.2216$~\cite{FerrenbergXuLandau}, i.e.\
$\tanh\beta_c=e^{-2K_c^{\rm Ising}}$. The mirror relation of
Sec.~\ref{supp:sm:fusion} places the same singularity at $s=0$:
\begin{equation}\label{supp:eq:Tc0}
T_c^{\rm proj}(s{=}1)=T_c^{\rm proj}(0)\approx1.3133 ,
\end{equation}
with three-dimensional Ising exponents (an analytic reparametrization of the coupling cannot alter the universality class).

The reversal of temperature ordering in this mapping has a direct loop-gas
interpretation. Because $K$ decreases with $\beta$, the
Ising-\emph{ordered} phase corresponds to the \emph{high}-temperature side of
the locally projected paramagnet. In the classical loop representation above
the transition is the proliferation of the closed loops of $x_l=-1$
links---electric strings---mirroring the magnetic flux-loop proliferation
that sets $T_c^{\rm proj}(0)$ at $s=0$, in agreement with the
$\rho_{\rm loc}$ mirror symmetry. Without $P_{\rm vertex}$, all contractible
charge sectors are summed with equal weight, the link qubits decouple, and
no transition survives; the singularity is entirely a property of the
locally projected ensemble $\rho_{\rm loc}$.

\subsubsection{Finite-\texorpdfstring{$\mu$}{mu} regularization}

A soft local projection through the star term $-\mu\sum_v S^{(1)}_v$, as in
our QMC, merely inserts $\tanh(\beta\mu)^{|S|}$ into the expansion above [use
$e^{\beta\mu S^{(1)}_v}=\cosh(\beta\mu)
[1+\tanh(\beta\mu)S^{(1)}_v]$], i.e.\ a uniform magnetic field
$h\coloneqq-\tfrac12\ln\tanh(\beta\mu)$ acting on the Ising spins. For any
finite $\mu$ this field is nonzero and rounds the thermodynamic Ising
singularity into a sharp crossover: finite $\mu$ is a regulator toward
$\rho_{\rm loc}$, not an exact projected ensemble. The exact projected
transition is recovered only in the local-projection limit $\mu\to\infty$
taken at fixed $L$ \emph{before} $L\to\infty$ (equivalently, along any
schedule with $h(L)\,L^{y_h}\to0$). Holding $\beta\mu=4$ fixed, as in
Sec.~\ref{supp:sm:proj-qmc}, gives the temperature-independent residual field
$h\approx3.4\times10^{-4}$ and rounding length
$\xi_h\sim h^{-1/y_h}\approx25$ ($y_h=2.48185$, the three-dimensional Ising
magnetic exponent~\cite{KPSDV}); the relevant scaling variable is then
$hL^{y_h}\approx0.16$ at $L=12$ and $0.33$ at $L=16$---small but
\emph{not} asymptotically zero. The finite-$\mu$ transition curve we report
is therefore a preasymptotic finite-regulator estimate of the $\mu\to\infty$
projected transition, not the exact-projection singularity itself. Two
distinct regulators appear in the numerics and must not be merged:
$\beta\mu=4$ held fixed (this section, $h$ temperature independent) and
$\mu=4$ held fixed (Sec.~\ref{supp:sm:tstar-scaling}, static charge cost $2\mu$,
$h$ temperature dependent). Both approach $P_{\rm loc}$ as
$\mu\to\infty$, and we compare across them only through observables for
which the residual field is negligible at the sizes used, never by combining
finite-$\mu$ data of different schedules. The same remarks apply to the
entire mirror arm $s>\tfrac12$ of the locally projected transition curve.

\subsection{Loop-gas representation of the two endpoints: winding sectors and degeneracy}
\label{supp:sm:s1-loops}
The two locally projected endpoints also share a \emph{loop-gas} representation that displays their winding-sector bookkeeping explicitly. Both representations are per species, equivalently per copy of the exactly decoupled dual pair; the $s=0$ gas below is the gauge member $\tilde H(0)$ of the KT pair, the two-species locally projected traces are the corresponding squares, and the six noncontractible charges give the full-model sector count $2^6=8\times8$ of Eq.~(\ref{supp:eq:Ploc-sectors}). Throughout, $w(\mathcal C)\in H_1(T^3;\mathbb Z_2)=\mathbb Z_2^3$ denotes the triple of winding parities of a link subset $\mathcal C$ around the three cycles of the periodic lattice.

At the $s=1$ endpoint, the classical representation derived above gives
\begin{equation}\label{supp:eq:loop-para}
Z_P^{(1)}(\beta)=e^{\beta N_l}\!\!\sum_{\mathcal C\ \mathrm{closed}}\!\!e^{-2\beta|\mathcal C|},
\end{equation}
where $\mathcal C$ runs over all even-degree link subsets of the \emph{direct} lattice---the full cycle space, with \emph{every} winding class $w(\mathcal C)$ admitted; the Gauss constraint does not distinguish a contractible loop from one winding around the torus.

At the $s=0$ endpoint, define
$Z_P^{(0)}(\beta)\coloneqq\mathrm{Tr}\,[e^{\beta\sum_p B_p}P_{\rm vertex}]$
and evaluate it in the $Z$ basis, where $e^{\beta\sum_p B_p}$ is diagonal and
only the two elements $S=\emptyset$ and $S=\mathcal V$ of the Gauss group
act trivially, $\langle z|P_{\rm vertex}|z\rangle=2^{1-N_v}$. In the
classical sum over $\{z_l\}$ the plaquette eigenvalues
$b_p\coloneqq\prod_{l\in\partial p}z_l$ define frustrated plaquettes when
$b_p=-1$; these live on links of the \emph{dual} (again simple-cubic)
lattice. The identity $b=\delta z$ in additive $\mathbb F_2$ notation
implies that the frustration set $\mathcal C^*$ is a closed dual-loop
configuration with zero total noncontractible winding parity in each
direction; it may contain contractible loops as well as pairs of
noncontractible loops. For fixed $b=\delta z$, its realizations form an
affine copy of $\ker\delta$. Exact $1$-cochains generated by vertex gauge
transformations number $2^{N_v-1}$ (the global transformation is
redundant), while $H^1(T^3;\mathbb Z_2)=\mathbb Z_2^3$ supplies the eight
noncontractible Wilson-line classes. Each admissible pattern is therefore
realized by exactly $2^{N_v-1}\times2^3$ configurations $\{z_l\}$. Hence
\begin{equation}\label{supp:eq:loop-ssb}
Z_P^{(0)}(\beta)=8\,e^{3\beta L^3}\!\!\sum_{\mathcal C^*:\,w(\mathcal C^*)=0}\!\!e^{-2\beta|\mathcal C^*|},
\end{equation}
the prefactor $8=2^3=|H_1(T^3;\mathbb Z_2)|$ being the topological ground-space degeneracy of the three-dimensional toric code (with all $S^{(1)}_v$ and $B_p$ fixed, the three winding Wilson-line classes remain). Here ``contractible'' refers to the combined loop configuration having zero total winding parity: a pair of parallel noncontractible flux loops is contractible as a combined configuration and is perfectly admissible at $s=0$; only \emph{odd} total winding is forbidden.

Despite their different winding restrictions, Eqs.~(\ref{supp:eq:loop-para}) and
(\ref{supp:eq:loop-ssb}) have the same bulk singularity. They use the same local
even-subgraph weights of the three-dimensional Ising high-temperature
expansion,
$Z_{\rm Ising}=2^{N_v}(\cosh K)^{N_l}
\sum_{\mathcal C}(\tanh K)^{|\mathcal C|}$, under the identification
$\tanh K=e^{-2\beta}$. The first sum contains all winding classes, whereas
the second contains the zero-total-winding class and carries an explicit
factor $8$. These differences shift $\ln Z$ only subextensively, as shown
below; their free-energy densities therefore have the same singularity at
$\tanh K_c^{\rm Ising}=e^{-2\beta_c}$, again Eq.~(\ref{supp:eq:Tc0}).

\subsubsection{Winding sectors, interface tension, and degeneracy}

Let $Z_w$ denote the loop-gas partition function restricted to winding class
$w$; then $Z=\sum_{w\in\mathbb Z_2^3}Z_w$. Weighting configurations by
$(-1)^{w_i(\mathcal C)}$ is, in the Ising dictionary, an antiperiodic twist
of the spin model, as follows. Fix a plane normal to direction $i$ and let
$B_i$ be the set of direction-$i$ bonds that cross it (one per transverse
column, $L^2$ in all). Reversing the coupling on those bonds, $K\to-K$ on
$B_i$, uses
$e^{-K\sigma_v\sigma_{v'}}=\cosh K\,
(1-\tanh K\,\sigma_v\sigma_{v'})$ and therefore inserts a factor
$\varepsilon_b\coloneqq-1$ per bond of $B_i$ in the expansion above; each loop
configuration then acquires
$\prod_{b\in\mathcal C}\varepsilon_b=(-1)^{|\mathcal C\cap B_i|}$,
\begin{equation}\label{supp:eq:ZAP}
Z^{\rm AP}_i=2^{N_v}(\cosh K)^{N_l}\!\!\sum_{\mathcal C\,\text{closed}}\!\!(-1)^{|\mathcal C\cap B_i|}(\tanh K)^{|\mathcal C|}.
\end{equation}
For a closed subgraph the number of crossings of the plane has the parity of the winding number, $|\mathcal C\cap B_i|\equiv w_i(\mathcal C)\ (\mathrm{mod}\ 2)$---contractible loops cross the plane an even number of times, direction-$i$ winding loops an odd number. Hence $(-1)^{|\mathcal C\cap B_i|}=(-1)^{w_i(\mathcal C)}$ and $Z^{\rm AP}_i=Z_{w_i=0}-Z_{w_i=1}$. That $Z^{\rm AP}_i$ is genuinely the antiperiodically continued partition function is seen by flipping $\sigma_v\to-\sigma_v$ on the entire half-space beyond the plane: this restores $+K$ on every bond of $B_i$ while turning the periodic identification around the $i$-cycle into an antiperiodic one. Dividing by $Z^{\rm P}=Z_{w_i=0}+Z_{w_i=1}$, the common high-temperature-expansion prefactor cancels. In the dense-loop (Ising-ordered) phase,
\begin{align}\label{supp:eq:twist}
\big\langle(-1)^{w_i(\mathcal C)}\big\rangle&=\frac{Z_{w_i=0}-Z_{w_i=1}}{Z_{w_i=0}+Z_{w_i=1}}=\frac{Z^{\rm AP}_i}{Z^{\rm P}}\nonumber\\
&=e^{-\sigma_i(K)\,L^2\,[1+o(1)]},
\end{align}
with $\sigma_i(K)$ the Ising interface tension at coupling $K$. In the \emph{dilute} phase (Ising disordered, $\sigma_i=0$ but winding loops carrying a line tension) one finds $Z_{w\neq0}/Z_0\sim e^{-cL}\to0$; a single sector dominates, the full-cycle and zero-winding sums agree up to exponentially small terms, and the $\ln8$ of Eq.~(\ref{supp:eq:loop-ssb}) is an innocuous constant. In the \emph{dense} phase (Ising ordered, $\sigma_i>0$) Eq.~(\ref{supp:eq:twist}) gives $\langle(-1)^{w_i(\mathcal C)}\rangle\to0$, i.e.\ $Z_w\to Z_0$ for every $w$; all $2^3=8$ winding sectors become degenerate, and $\sum_wZ_w\simeq8\,Z_0$ reproduces the explicit factor $8$ of the $s=0$ representation. The spontaneous breaking of the $0$-form $\mathbb Z_2$ in spin variables thus appears in the dual loop variables as equidistribution over the $\mathbb Z_2$ winding sectors. Correspondingly, the local order parameter maps to open defect lines: $\langle\sigma_0\sigma_r\rangle$ is the loop sum with one open string from $0$ to $r$, and long-range order is the vanishing of its line tension.

At low temperature the $s=0$ endpoint carries the factor $8$ explicitly as the thermally stable topological degeneracy of the deconfined dome, whereas the $s=1$ loop gas is dilute and sector-selected. At high temperature the $s=1$ gas is dense and its eight winding sectors are degenerate, matching the explicit factor on the other side. This redistribution changes finite-size degeneracies, but not the bulk free-energy density, as allowed by the non-invertible duality.

\section{Finite-regulator phase-diagram calculation}
\label{supp:sm:numerics}

\subsection{Finite-\texorpdfstring{$\mu$}{mu} regulation toward
\texorpdfstring{$\rho_{\rm loc}$}{rho-loc} and the measured transition curve}
\label{supp:sm:proj-qmc}
The data of Fig.~1(a) of the Letter are obtained with a finite-$\mu$
regulator toward the locally projected state $\rho_{\rm loc}$: the
contractible generators are favored energetically through the soft terms
$-\mu(\sum_vS^{(1)}_v+\sum_cS^{(1)}_c)$, with $\beta\mu=4$ held fixed along
every temperature scan. By the finite-$\mu$ mapping in
Sec.~\ref{supp:sm:s1-ising}, the residual
Ising field $h=-\tfrac12\ln\tanh(\beta\mu)\approx3.4\times10^{-4}$ is
temperature independent and small at our sizes, but it is nonzero; these
simulations are therefore preasymptotic estimates of $\rho_{\rm loc}$, not
exact projected traces. The crossover-based arm estimates in Fig.~1(a) are measured
by discrete-time worldline QMC applied to $H(s)$ in the cluster-stabilizer
frame, including both soft terms. The supplementary wall-height estimates of
this section and Sec.~\ref{supp:sm:tstar-scaling} (Figs.~\ref{supp:fig:pdDT} and
\ref{supp:fig:proj-qmc}) instead use continuous-time QMC of the decoupled gauge
copy in the KT-image frame, again with a finite soft constraint
(ParaToric~\cite{ParaToric}).

The KT identity, Eq.~(\ref{supp:eq:proj-identity}), applies at finite $L$ only to
$P_+$. After exact local projection, Eq.~(\ref{supp:eq:sector-bound}) equates its
bulk free-energy density with that of $\rho_{\rm loc}$. The finite-$\mu$
simulations approach $\rho_{\rm loc}$ only in the order of limits stated
there and are therefore not identified with $P_+$ at finite $L$.  The
present arm campaign does not include a systematic $\mu\to\infty$
extrapolation; its coordinates are consequently reported only as
finite-regulator estimates.

\subsubsection{Continuous-transition arms}

Let
$G\coloneqq\sum_vS_v^{(1)}+\sum_cS_c^{(1)}$.  Along every scan
$B\coloneqq\beta\mu=4$ is fixed, and the sampled exponent is
$-\beta H(s)+BG$.  The energy conjugate to $\beta$ on this path is therefore
the physical energy $E_{\rm phys}\coloneqq\langle H(s)\rangle$, not the energy
$E_\mu\coloneqq\langle H(s)-\mu G\rangle$ of a fixed-$\mu$ Hamiltonian.  In
the measured ensemble these quantities satisfy
\begin{equation}\label{supp:eq:Ephys-arm}
E_{\rm phys}=E_\mu+\mu\langle G\rangle,\qquad \mu=4T .
\end{equation}
At each $s$ on the grid $\Delta s=0.05$, we locate the maximum slope of the
seed-averaged $E_{\rm phys}(T)$ for $L=6,8,10,12,16$ ($4000$
thermalization and $6000$--$16000$ measurement sweeps per size, $32$ seeds).
Denote this maximum-slope temperature by $T_{\rm inf}^{(\mu)}(s,L)$ and its
thermodynamic extrapolation along the arm by $T_{\rm arm}^{(\mu)}(s)$.  We
fit the finite-size locations to
\begin{equation}\label{supp:eq:arm-fss}
T_{\rm inf}^{(\mu)}(s,L)
=T_{\rm arm}^{(\mu)}(s)+A L^{-1/\nu},\qquad \nu=0.630 .
\end{equation}
Here $A$ is a nonuniversal amplitude and $\nu=0.630$ is the
three-dimensional Ising correlation-length exponent.  The fits use
$L=8,10,12,16$ and bootstrap the $32$ independent seeds at each
point.  Their unanchored endpoint value is $1.337$, rather than the exact
projected value $1.3133$; this difference measures the combined finite-field,
finite-grid, and discretization bias of this procedure.

We also monitor
\begin{equation}\label{supp:eq:Rmu-arm}
R_\mu\coloneqq\frac{\beta^2\operatorname{Var}(E_\mu)
 -\langle N_{\rm off}\rangle}{N},\qquad N=6L^3,
\end{equation}
where $N_{\rm off}$ is the total number of off-diagonal worldline events of
all three update types.  This
quantity is a useful energy-fluctuation diagnostic for the simulated
discrete-time action, but it is \emph{not} the heat capacity
$d\langle H\rangle/dT$ along fixed $B$.  The red circles in Fig.~1(a) of the
Letter are the peak-and-drift crossover estimates from $R_\mu$; they give $1.320(6)$ at
the endpoint, within about one finite-size-scaling error bar of the exact
$T_c^{\rm proj}(0)=1.3133$.  After the physical-energy analysis is anchored only by its
endpoint offset, its interior locations differ from the displayed ones by at
most $0.016$ on the measured grid.  We regard this method-to-method spread,
rather than the much smaller fit errors alone, as the relevant numerical
systematic.  Accordingly the red points give the shape of the
finite-regulator arm and should not be read as precision determinations of
the $\mu\to\infty$, $\Delta\tau\to0$ transition curve, where $\Delta\tau$
is the imaginary-time step.  The exact mirror
symmetry is imposed after the one-sided analysis.  A dimensionless Binder or
correlation-length crossing, together with explicit regulator and Trotter
scaling, would be needed for precision critical coordinates and an
independent universality determination.

\subsubsection{First-order wall}

Across the self-dual line we scan $s$ at fixed
temperature on the fine grid $\Delta s=0.01$ for
$0.02\le T\le0.44$, initializing each run in deconfined- and
confined-favored configurations. The order parameter
$\mathcal M\coloneqq\partial_sH$ of the Letter is measured, in the gauge frame of the
simulated copy, as $\langle B_p\rangle-\langle X_l\rangle$ per link; since $H(s)$
is link--plaquette symmetric, both species share this per-site density; hence
$\langle\mathcal M\rangle/N
=\langle B_p\rangle-\langle X_l\rangle$ with $N=6L^3$. It is,
by the finite-temperature Hellmann--Feynman relation, the isothermal slope
of the finite-$\mu$ free energy per site.  At low temperature the two
preparations remain on distinct branches during the run.  We count a branch
as stable only if its signed density exceeds $0.12$ and the first- to
last-quarter drift is below $0.12$; every seed of both preparations must pass
these tests.  This operational criterion distinguishes persistent
two-branch data from the kinetic splitting left after one branch relaxes.
The branch-symmetric combination locates the regulated wall at
$|s_c-\tfrac12|\le0.01$ wherever the criterion is met, and the branch values
obey the approximate duality antisymmetry
$|\langle\mathcal M\rangle_{\rm dec}(s)
+\langle\mathcal M\rangle_{\rm conf}(1-s)|/N\le0.02$.
Exact pinning refers to the target projected states discussed in
Sec.~\ref{supp:sm:fusion}; agreement of the finite-$\mu$ data with it is a
regulator check.

\subsubsection{Critical endpoint}

\begin{figure}[!t]
\centering
\includegraphics[width=0.9\columnwidth]{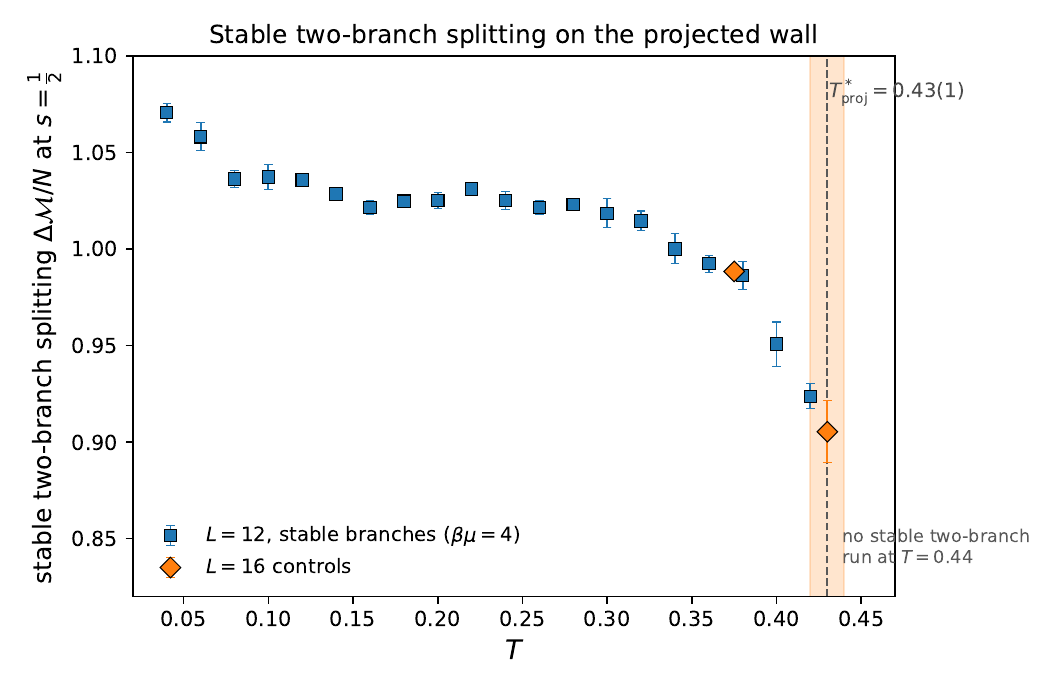}
\caption{Stable two-branch splitting
$\Delta\mathcal M/N\,(T)$ of the duality-odd density across the
finite-$\mu$ wall, used as a regulator estimate for $\rho_{\rm loc}$.  Only
points for which both prepared branches pass the stability criterion in the
text are shown.  The splitting stays large ($\approx0.9$--$1.0$) through
$T=0.43$ and no stable two-branch run is found at $0.44$; its thermodynamic
collapse is not resolved.  The finite-regulator estimate
$T^{*}_{\rm proj}=0.43(1)$ (star in Fig.~1(a) of the Letter) combines the
observed loss of two-branch stability between $0.43$ and $0.44$ with the independent
finite-size scaled-field crossing, and does not use an extrapolation of this
splitting or of the continuous arms.}
\label{supp:fig:pdDT}
\end{figure}

Two finite-regulator indicators bracket
the wall top.  First, the hysteresis strips show a resolved two-branch splitting
through $T=0.42$; an $L=16$ control remains two-branched at $T=0.43$, while
no stable two-branch run is found at $T=0.44$.  The measured splitting
$\Delta\mathcal M/N\,(T)$ (Fig.~\ref{supp:fig:pdDT}) remains
$\approx0.9$--$1.0$ up to the last stable two-branch temperature, but its
collapse is not resolved closely enough to support a power-law
extrapolation.  Second, the scaled-field crossing of
Sec.~\ref{supp:sm:tstar-scaling} gives $0.43(1)$.  We do not use an extrapolation
of either the jump or the continuous-arm points to determine the endpoint.
Together these two criteria give the finite-regulator estimate
\begin{equation}
T^{*}_{\rm proj}=0.43(1).
\end{equation}
The exact mirror symmetry pins the projected wall to $s=\tfrac12$, but this
campaign does not extrapolate its height in $\mu$.  The value above is
therefore evidence for the endpoint of the target $\rho_{\rm loc}$ diagram,
not a determination after taking the local-projection limit.  The phase
topology inferred from reflection places the low-temperature coexistence
line below this endpoint and the intervening disordered phase between the
two arms above it.  The arm data stop short of the junction and do not
independently determine where the three boundaries meet.

\subsection{Scaling-field estimate of the wall height at the self-dual point}
\label{supp:sm:tstar-scaling}
\begin{figure}[!t]
\centering
\includegraphics[width=0.9\columnwidth]{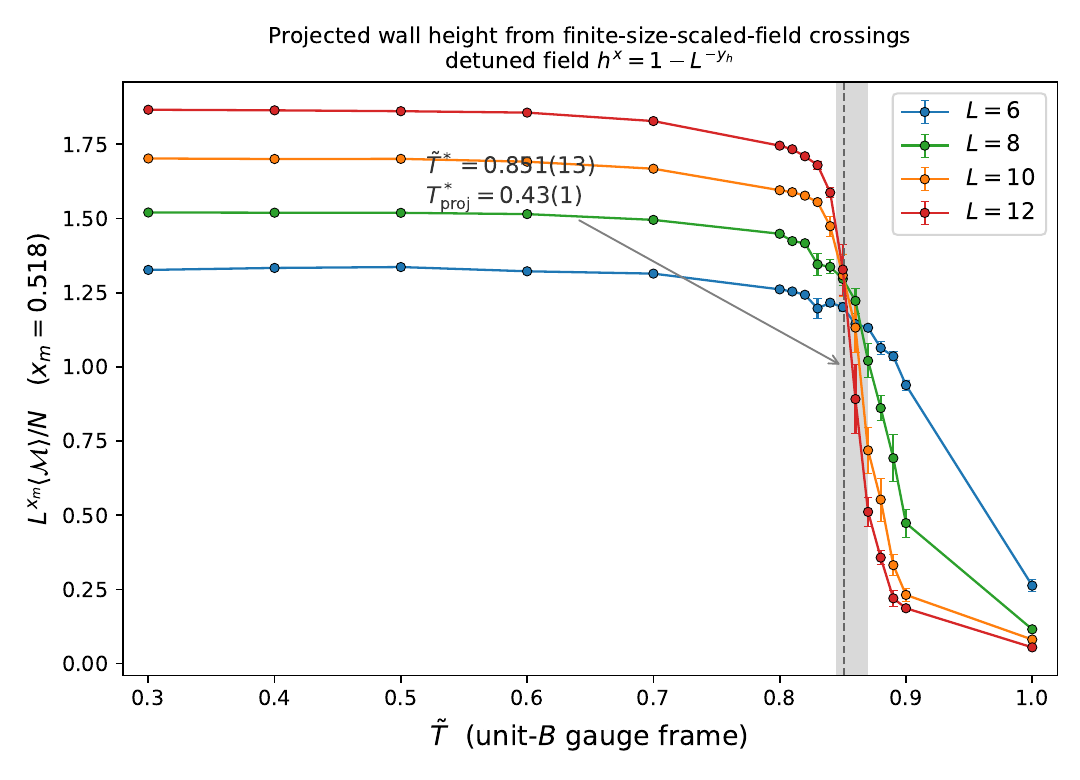}
\caption{Finite-$\mu$ regulator estimate of the $\rho_{\rm loc}$ wall height
from scaled-field crossings. The scaled duality-odd density
$L^{x_m}\langle\mathcal M\rangle/N$ at the detuned field
$h^x=1-L^{-y_h}$ (three-dimensional Ising exponents) is size independent at
the endpoint; the curves for $L=6$--$12$ cross in a narrow window (shaded),
giving the finite-regulator estimate in Eq.~(\ref{supp:eq:TstarQMC}).}
\label{supp:fig:proj-qmc}
\end{figure}

An independent estimate of the wall top uses the fixed soft-penalty strength
$\mu=4$ (static contractible-charge cost $2\mu=8$, single-defect Boltzmann
factor below $e^{-8/\tilde T}\lesssim10^{-3}$ over the temperatures studied;
here $\tilde T$ is the temperature of the unit-coupling gauge Hamiltonian
$-\sum_pB_p-h^x\sum_lX_l$, related to the frame of the Letter by
$T=(1-s)\tilde T$). The frame is that of the KT image; the identical protocol
in the $H(s)$ frame detunes $s=\tfrac12-\delta_L/4$
[$h^x\coloneqq s/(1-s)$, $dh^x/ds|_{1/2}=4$], and its repetition is part of the direct
re-anchoring campaign of Sec.~\ref{supp:sm:proj-qmc}. At $h^x=1$, the KT trace
identity is exactly self-dual for the all-$+$ full-fusion projector $P_+$.
Equation~(\ref{supp:eq:sector-bound}) transfers the thermodynamic wall position to
$\rho_{\rm loc}$ while allowing different finite-$L$ distributions. The
fixed-$\mu$ ensemble simulated here has no exact projected-ensemble identity;
its near-symmetry therefore serves only as a convergence check. In the target projected states
the duality-odd density
$\langle\mathcal M\rangle/N=\langle B_p\rangle-\langle X_l\rangle$
is the order parameter of the wall and, by the finite-temperature
Hellmann--Feynman relation, the isothermal free-energy slope conjugate to the
coupling; its two-sided jump is the latent slope of the first-order line. We
estimate the finite-regulator wall height by finite-size-scaled-field crossings. At
each size the field is detuned by $\delta_L\coloneqq L^{-y_h}$ on the
deconfined-favored side, $h^x=1-\delta_L$, with the three-dimensional Ising
magnetic exponent $y_h=2.48185$~\cite{KPSDV} (the endpoint of a symmetric
first-order line is an Ising critical point in the
$\mathcal M$ channel), and the scaled order parameter
$L^{x_m}\langle\mathcal M\rangle/N$,
$x_m\coloneqq3-y_h=0.51815$, is size
independent at the endpoint, increasing with $L$ below it and decreasing
above. Because the detuned field favors the phase in which the chains are
prepared, this protocol reduces initialization trapping. With
$L=6,8,10,12$ ($2$--$2.5\times10^5$ samples spaced $2L^3$ updates, $6$ seeds
on the coarse $\Delta\tilde T=0.10$ grid and $8$ on the fine
$\Delta\tilde T=0.01$ grid that resolves the crossing), the pair crossings sit
at $\tilde T_\times=0.864\,(6,8)$, $0.851\,(8,10)$,
$0.851\,(10,12)$; the drift saturates by the two larger-$L$ pairs
(correction-to-scaling exponent $\omega=0.83$~\cite{KPSDV}), giving the
regulator estimate $\tilde T^*=0.851(13)$, i.e.
\begin{equation}\label{supp:eq:TstarQMC}
T^*_{\rm proj}=0.43(1)\;=\;0.33(1)\,T_c^{\rm proj}(0),
\end{equation}
where the quoted uncertainty is a conservative envelope of the pair-crossing
spread and the finite scaling-field amplitude, rather than a bootstrap error
or a controlled correction-to-scaling extrapolation
(Fig.~\ref{supp:fig:proj-qmc}).  It is consistent with the two-branch bracket of
Sec.~\ref{supp:sm:proj-qmc}.

A further check supports Eq.~(\ref{supp:eq:TstarQMC}). Long single-chain runs at
$h^x=1$ test how closely the finite-$\mu$ histogram of
$\mathcal M$ approaches the symmetric histogram of the target
projected state. They show genuine two-phase tunneling only at low
temperatures. At $T=0.45$ the equilibrated distribution has collapsed:
$\langle|\mathcal M|\rangle/N$ falls with system size as a
fluctuation ($0.40,0.28,0.14$ for $L=4,6,8$, $16$ seeds) rather than converging to a
finite coexistence value. This collapse places the end of the regulated wall
below $T=0.45$.

\section{Twisted membrane index and exact boundary lines}
\label{supp:sm:index-theory}

\subsection{Twisted membranes in \texorpdfstring{$\rho_{\rm loc}$}{rho-loc}: exact fixed-point evaluation}
\label{supp:sm:s1-twist}
The same machinery evaluates the twisted membrane of the Letter in
$\rho_{\rm loc}$. We work at the two fixed points of the RBH interpolation,
with plaquette and link qubits, cluster stabilizers $K_p\coloneqq X_pB_p$ and
$K_l\coloneqq X_lB_l$, and the paramagnet
$H_{\rm para}\coloneqq-\sum_pX_p-\sum_lX_l$. The strong symmetry under $G_{\rm loc}$
is generated by the contractible one-form operators $S^{(1)}_c$ (cubes) and
$S^{(1)}_v$ (vertices), with projector
$P_{\rm loc}\coloneqq\prod_c\frac{1+S^{(1)}_c}{2}
\prod_v\frac{1+S^{(1)}_v}{2}$; the noncontractible charges remain summed.
The twist is the flat sign cocycle of the Letter, $\eta_p\coloneqq-1$ exactly on
the plaquettes pierced by the flux loop $\gamma$ and $+1$ elsewhere
($B_p\to\eta_pB_p$; flat,
$\prod_{p\in\partial c}\eta_p=1$, but not exact, its class
Poincar\'e-dual to $[\gamma]$). Changing its representative by an exact
cocycle is a local change of variables; consequently, only the class of
$\gamma$ enters.

We first evaluate the twist in the fixed-point loop gas. In the joint
eigenbasis (stabilizers on the cluster side, $X$'s on the paramagnet side;
note $S^{(1)}_c=\prod_{p\in\partial c}K_p$ since
$\prod_{p\in\partial c}B_p=I$) the projector constrains the flipped
plaquette variables identically on both sides. Faces map to links of the dual
lattice, cubes to dual vertices, and the cube constraint is the dual
vertex-star condition---the flipped set is a configuration of
\emph{closed loops $\mathcal C^*$ on the dual lattice}, with relative weight
$e^{-2\beta|\mathcal C^*|}$. Unlike the zero-winding KT gauge sum in
Eq.~(\ref{supp:eq:loop-ssb}), this fixed-point cluster/paramagnet loop gas sums all
winding classes. For a closed surface $\Sigma$ the measured sign
$(-1)^{|\mathcal C^*\cap\Sigma|}$ pairs the $1$-cycle $\mathcal C^*$ with
the dual $1$-cocycle of $\Sigma$; therefore, contractible defect loops drop
out and only winding defects count. On the cluster side the twist multiplies
$K_p\to\eta_pK_p$ with a flat cocycle
($\prod_{p\in\partial c}\eta_p=1$), and the change of variables
$k_p\to\eta_pk_p$ [$k_p=\pm1$ the $K_p$ eigenvalue] pulls out
$\prod_{p\in\Sigma}\eta_p=(-1)^{\Sigma\cdot\gamma}$; on the paramagnet side
the twist acts only through the $Z$-dressings and is inert. It follows that
\begin{equation}\label{supp:eq:tw-exact}
\langle S_\Sigma^{(1)}\rangle_{\rm tw}^{P_{\rm loc}}=
\begin{cases}
(-1)^{\Sigma\cdot\gamma}\,\big\langle(-1)^{w(\mathcal C^*)\cdot[\Sigma]}\big\rangle_{\rm loop} & (H_{\rm SPT}),\\[2pt]
\phantom{(-1)^{\Sigma\cdot\gamma}\,}\big\langle(-1)^{w(\mathcal C^*)\cdot[\Sigma]}\big\rangle_{\rm loop} & (H_{\rm para}).
\end{cases}
\end{equation}

\subsubsection{Twisted reflection throughout the interpolation}

The fixed-point argument extends to the whole interpolation. Conjugating the
flux-threaded interpolation $H_{\rm tw}(s)$ by $U$ moves the cocycle from the
$X_pB_p$ terms onto the $X_p$ terms; the further conjugation by
$W\coloneqq\prod_{p:\,\eta_p=-1}Z_p$---which commutes with $P_{\rm loc}$ by
the flatness of $\eta$ and obeys
$WS_\Sigma^{(1)}W^\dagger=(-1)^{\Sigma\cdot\gamma}S_\Sigma^{(1)}$---removes
it there and reinstates it on the $X_pB_p$ terms, yielding
$(WU)H_{\rm tw}(s)(WU)^\dagger=H_{\rm tw}(1-s)$ exactly. Hence, at every
$(s,T)$ and finite $L$,
\begin{equation}\label{supp:eq:tw-reflection}
\langle S_\Sigma^{(1)}\rangle_{\rm tw}^{P_{\rm loc}}(s,T)=(-1)^{\Sigma\cdot\gamma}\,\langle S_\Sigma^{(1)}\rangle_{\rm tw}^{P_{\rm loc}}(1-s,T),
\end{equation}
the finite-volume twisted-reflection identity; at the fixed points it reduces to Eq.~(\ref{supp:eq:tw-exact}).

\subsubsection{Sector free-energy representation of \texorpdfstring{$F_\Sigma$}{F Sigma}}

A single conjugation turns the twisted membrane into an ordinary two-sector
free-energy difference, which at the fixed points reduces to standard Ising
line and interface free energies. The operator $W$ defined above commutes
with $P_{\rm loc}$ (flatness of $\eta$) and obeys
$WS_\Sigma^{(1)}W^\dagger=(-1)^{\Sigma\cdot\gamma}S_\Sigma^{(1)}$ and
$WH_{\rm tw}(s)W^\dagger=H(s)+2s\sum_{p:\,\eta_p=-1}X_p$; it trades the flux
line for a line of \emph{reversed} transverse field. Inserting
$W^\dagger W=I$ under the trace,
\begin{equation}\label{supp:eq:FSigma-def}
\langle S_\Sigma^{(1)}\rangle_{\rm tw}^{P_{\rm loc}}=(-1)^{\Sigma\cdot\gamma}\,\frac{Z_{\rm even}-Z_{\rm odd}}{Z_{\rm even}+Z_{\rm odd}}=(-1)^{\Sigma\cdot\gamma}\tanh\frac{F_\Sigma}{2},
\end{equation}
where $F_{\rm even/odd}\coloneqq-\ln Z_{\rm even/odd}$ are the dimensionless free energies of this defect model in the two sectors of the conserved crossing parity $m_\Sigma\coloneqq\prod_{p\in\Sigma}k_p=(-1)^{w(\mathcal C^*)\cdot[\Sigma]}$ (an even/odd number of flux loops threading $\Sigma$), and $F_\Sigma\coloneqq F_{\rm odd}-F_{\rm even}$. Equivalently, if $\mathfrak F_{\rm even/odd}\coloneqq-T\ln Z_{\rm even/odd}$ are the free energies in energy units, then $F_\Sigma=\beta(\mathfrak F_{\rm odd}-\mathfrak F_{\rm even})$. Every local move preserves $m_\Sigma$; hence Eq.~(\ref{supp:eq:FSigma-def}) holds at every $s$. At the fixed points the defect model is the loop gas of Eq.~(\ref{supp:eq:tw-exact}) and $F_\Sigma$ follows analytically.

Below $T_c^{\rm proj}(0)$ the dual loops are dilute, with fugacity
$u\coloneqq e^{-2\beta}$ per unit length. The cheapest odd configuration is a
\emph{single} loop winding once around the cycle dual to $\Sigma$; advancing
it layer by layer is governed by the directed-loop transfer matrix
$\mathcal T=u\,I+u^2A_\perp+O(u^3)$ ($I$ the identity; a straight step costs
$u$, a unit transverse step $u^2$; $A_\perp$ the transverse adjacency,
coordination $z_\perp\coloneqq4$), whose top eigenvalue
$u(1+z_\perp u+\cdots)$ sets the winding line tension
\begin{equation}\label{supp:eq:linetension}
\kappa(\beta)=2\beta-4e^{-2\beta}+O(e^{-4\beta})>0 .
\end{equation}
This low-fugacity expansion establishes $\kappa>0$ directly near $T=0$.
Throughout the full open region $0<T<T_c^{\rm proj}(0)$, positivity follows
instead from exponential clustering in the disordered three-dimensional
Ising phase~\cite{AizenmanBarskyFernandez}. Under the Ising loop/gauge
dictionary~\cite{Wegner}, its finite inverse correlation length is the
positive free-energy cost per unit length of a winding loop. Thus
$Z_{\rm odd}/Z_{\rm even}=\mathrm{poly}(L)e^{-\kappa L}$ for every fixed
$T<T_c^{\rm proj}(0)$, with $\kappa$ vanishing only on approaching the
critical point. It follows that
$F_\Sigma=\kappa(\beta)L[1+o(1)]\to\infty$ and
$\langle S_\Sigma^{(1)}\rangle_{\rm tw}^{P_{\rm loc}}
\to(-1)^{\Sigma\cdot\gamma}$, with corrections
$O(\mathrm{poly}(L)e^{-\kappa L})$.

Above $T_c^{\rm proj}(0)$ the loops condense (the Ising-ordered phase of
Sec.~\ref{supp:sm:s1-loops}), and the odd sector differs from the even one by an
antiperiodic seam across $\Sigma$---an interface of area $L^2$. By
Eq.~(\ref{supp:eq:twist}),
$(Z_{\rm even}-Z_{\rm odd})/(Z_{\rm even}+Z_{\rm odd})
=Z^{\rm AP}/Z^{\rm P}=e^{-\sigma(\beta)L^2[1+o(1)]}$ with $\sigma$ the
strictly positive Ising interface tension throughout the ordered
phase~\cite{LebowitzPfister}. Hence,
\begin{equation}\label{supp:eq:FSigma-high}
F_\Sigma=2\,\mathrm{artanh}\frac{Z^{\rm AP}}{Z^{\rm P}}=2\,e^{-\sigma(\beta)L^2[1+o(1)]}\to0,
\end{equation}
and $\langle S_\Sigma^{(1)}\rangle_{\rm tw}^{P_{\rm loc}}\to0$. The two regimes meet at the loop-condensation point $\tanh K_c=e^{-2\beta_c}$, i.e.\ $T_c^{\rm proj}(0)$: $F_\Sigma$ crosses over from extensive in $L$ (a line tension) to exponentially small in $L$ (an interface), separating the two asymptotic forms of the membrane response.

\subsubsection{Unconditional fixed-point lines}
\label{supp:sm:s0-line}

Write
$I_L(s,T)\coloneqq
\langle S_\Sigma^{(1)}\rangle_{\rm tw}^{P_{\rm loc}}$
as in the Letter, and, whenever the thermodynamic limit exists, define
$\mathcal I(s,T)\coloneqq\lim_{L\to\infty}I_L(s,T)$. For
$T<T_c^{\rm proj}(0)\approx1.3133$ the fixed-point loop gas is dilute, winding
loops are suppressed by a line tension $\kappa>0$, and
$\langle(-1)^{w(\mathcal C^*)\cdot[\Sigma]}\rangle
=1-O(\operatorname{poly}(L)e^{-\kappa L})$; for
$T>T_c^{\rm proj}(0)$ it collapses with the
interface area law of Eq.~(\ref{supp:eq:twist}). Therefore, at finite $L$, the
two endpoints of the locally projected family obey unconditionally
\begin{equation}\label{supp:eq:fixedpoint-ratio-sm}
\begin{aligned}
\langle S_\Sigma^{(1)}\rangle_{\rm tw}^{P_{\rm loc}}
&=\frac{Z_{\rm Ising}^{\rm AP}}{Z_{\rm Ising}^{\rm P}}
\begin{cases}
(-1)^{\Sigma\cdot\gamma} & (s=0),\\
1 & (s=1),
\end{cases}\\[-2pt]
\tanh K&=e^{-2\beta}.
\end{aligned}
\end{equation}
Consequently,
\begin{equation}
\langle S_\Sigma^{(1)}\rangle_{\rm tw}^{P_{\rm loc}}\;\xrightarrow{L\to\infty}\;
\begin{cases}
(-1)^{\Sigma\cdot\gamma} & (s=0),\\
+1 & (s=1),
\end{cases}
\end{equation}
for all $0\le T<T_c^{\rm proj}(0)$, and both collapse to zero above
$T_c^{\rm proj}(0)$. At criticality the ratio in
Eq.~(\ref{supp:eq:fixedpoint-ratio-sm}) may approach an $O(1)$ constant;
quantization refers to open phases.  Writing $|\Sigma|\coloneqq L^2$ for the number
of plaquettes, or area, of the chosen membrane, the corresponding values
without $P_{\rm loc}$ are
$(-1)^{\Sigma\cdot\gamma}\tanh^{|\Sigma|}\beta$ at $s=0$ and
$+\tanh^{|\Sigma|}\beta$ at $s=1$; both vanish at every
$T>0$~\cite{Companion}. Away from the self-dual line, finite-temperature
interior phase values require the separate conditional argument below.

\subsubsection{Unconditional result on the self-dual
\texorpdfstring{$s=\tfrac12$}{s=1/2} line}
\label{supp:sm:selfdual-index-line}

For the defining odd intersection
$(-1)^{\Sigma\cdot\gamma}=-1$, the finite-volume reflection identity
Eq.~(\ref{supp:eq:tw-reflection}) gives
\begin{equation}\label{supp:eq:selfdual-index-zero}
I_L(\tfrac12,T)=-I_L(\tfrac12,T)=0 .
\end{equation}
This holds at every temperature and finite $L$, without a thermodynamic
limit or Hypothesis~(H).  Below the endpoint of the first-order line it is
the symmetry-enforced value of the canonical mixture on the coexistence
line, not a phase value assigned to either one-sided limit.

\subsubsection{Unconditional result on the \texorpdfstring{$T=0$}{T=0} line}
\label{supp:sm:T0-line}

At zero temperature the whole interpolation can be treated without
Hypothesis~(H).  We first consider $s'>\tfrac12$ at finite $L$.  Let
$\ket{\mathbf z}$ be a computational-basis state and
$\mathbf z^{(i)}$ the configuration obtained by flipping qubit $i$.  The
only two terms that connect these configurations are the paramagnetic and
cluster terms on that qubit.  Define
$c_l(\mathbf z)\coloneqq B_l(\mathbf z)$ for a link qubit and
$c_p(\mathbf z)\coloneqq\eta_pB_p(\mathbf z)$ for a plaquette qubit.  Then
\begin{equation}\label{supp:eq:PF-matrix-element}
\bra{\mathbf z^{(i)}}H_{\rm tw}(s')\ket{\mathbf z}
=-\big[s'+(1-s')c_i(\mathbf z)\big].
\end{equation}
Since $c_i(\mathbf z)=\pm1$, every one-qubit-flip matrix element is strictly
negative:
\begin{equation}
s'+(1-s')c_i(\mathbf z)\ge 2s'-1>0 .
\end{equation}
The one-qubit flips connect the full computational basis.  Thus
all diagonal entries vanish and $-H_{\rm tw}(s')$ is an irreducible
nonnegative matrix.  The
Perron--Frobenius theorem gives a unique ground state
$\ket{\psi_{s'}}$ whose computational-basis amplitudes are all strictly
positive.

Every generator of $G_{\rm loc}$ and the noncontractible membrane
$S_\Sigma^{(1)}$ commute with $H_{\rm tw}(s')$. The closedness identities
$\prod_{p\in\partial c}B_p=\prod_{l\ni v}B_l
=\prod_{p\in\Sigma}B_p=I$ cancel all $B$ dressings, leaving the pure-$X$ products
$S_c^{(1)}$, $S_v^{(1)}$, and $S_\Sigma^{(1)}$; in the computational basis
they are therefore sign-free permutation matrices, with every nonzero matrix
element equal to $+1$. Uniqueness makes $\ket{\psi_{s'}}$ an eigenstate of
each operator. Its strict positivity excludes eigenvalue $-1$, so
\begin{equation}
P_{\rm loc}\ket{\psi_{s'}}=\ket{\psi_{s'}},\qquad
S_\Sigma^{(1)}\ket{\psi_{s'}}=\ket{\psi_{s'}} .
\end{equation}
The projected zero-temperature state therefore selects this same ground
state and
\begin{equation}\label{supp:eq:T0-positive-half}
I_L(s',0)\coloneqq\lim_{\beta\to\infty}I_L(s',T)=+1,
\qquad \tfrac12<s'\le1 .
\end{equation}

For $0\le s<\tfrac12$, set $s'\coloneqq1-s$ and use the unitary
$Q\coloneqq WU$ of Eq.~(\ref{supp:eq:tw-reflection}).  It obeys
\begin{equation}
\begin{aligned}
QH_{\rm tw}(s)Q^\dagger&=H_{\rm tw}(s'),\\
QP_{\rm loc}Q^\dagger&=P_{\rm loc},\\
QS_\Sigma^{(1)}Q^\dagger
&=(-1)^{\Sigma\cdot\gamma}S_\Sigma^{(1)} .
\end{aligned}
\end{equation}
Consequently $Q^\dagger\ket{\psi_{s'}}$ is the unique ground state at $s$,
belongs to the image of $P_{\rm loc}$, and has membrane eigenvalue
$(-1)^{\Sigma\cdot\gamma}$.  For every finite $L$,
\begin{equation}\label{supp:eq:T0-line-exact}
I_L(s,0)\coloneqq\lim_{\beta\to\infty}I_L(s,T)
=(-1)^{\Sigma\cdot\gamma},\qquad 0\le s<\tfrac12 .
\end{equation}
For the defining odd intersection this is $-1$.  At $s=\tfrac12$ the exact
zero in Eq.~(\ref{supp:eq:selfdual-index-zero}) applies.  This twisted-sector
statement does not rely on the order or even the existence of a bulk
transition in the untwisted Hamiltonian.

Equation~(\ref{supp:eq:T0-line-exact}) uses the standard zero-temperature order
of limits
\begin{equation}
\mathcal I(s,0)\coloneqq
\lim_{L\to\infty}I_L(s,0)
=\lim_{L\to\infty}\lim_{\beta\to\infty}I_L(s,T).
\end{equation}
Interchanging the two limits, equivalently taking $T\downarrow0$ after the
finite-temperature thermodynamic limit, would require a bound uniform in
$L$ and is not implied by the finite-volume Perron--Frobenius argument.

\subsubsection{Unconditional result on the
\texorpdfstring{$T=\infty$}{T=infinity} line}
\label{supp:sm:Tinf-line}

At $\beta=0$ the Hamiltonian and its twist drop out before any
thermodynamic limit is taken. Conjugation by $W$, which preserves
$P_{\rm loc}$, supplies the same $(-1)^{\Sigma\cdot\gamma}$ prefactor as in
Eq.~(\ref{supp:eq:FSigma-def}). Using the group-average representation in
Eq.~(\ref{supp:eq:ensemble-glossary}),
\begin{equation}\label{supp:eq:Tinf-trace}
\begin{aligned}
I_L(s,\infty)
&=(-1)^{\Sigma\cdot\gamma}
\frac{\Tr[S_\Sigma^{(1)}P_{\rm loc}]}{\Tr P_{\rm loc}}\\[-2pt]
&=(-1)^{\Sigma\cdot\gamma}
\frac{\sum_{g\in G_{\rm loc}}\Tr[S_\Sigma^{(1)}g]}
{\sum_{g\in G_{\rm loc}}\Tr g}=0 .
\end{aligned}
\end{equation}
Indeed, $S_\Sigma^{(1)}$ is noncontractible and therefore does not belong to
$G_{\rm loc}$; $S_\Sigma^{(1)}g$ is a nonidentity Pauli operator for every
$g\in G_{\rm loc}$ and has zero trace.  The denominator is
$|G_{\rm loc}|\Tr P_{\rm loc}>0$.  Equivalently, the two membrane sectors
have equal rank,
$Z_{\rm even}=Z_{\rm odd}=\tfrac12\Tr P_{\rm loc}$, and hence $F_\Sigma=0$.
Thus
\begin{equation}\label{supp:eq:Tinf-line-exact}
I_L(s,\infty)=0
\end{equation}
for every $s$ and finite $L$, and therefore
$\mathcal I(s,\infty)=0$, without Hypothesis~(H).
For a soft constraint this exact projected state requires the projection
limit to be taken before $\beta=0$; fixed $\mu$ followed by $T\to\infty$
defines a different ensemble.

Collecting the four unconditional boundary statements for
$\Sigma\cdot\gamma=1$ gives
\begin{align}
\mathcal I(0,T)&=
\begin{cases}
-1,&0\le T<T_c^{\rm proj}(0),\\
0,&T>T_c^{\rm proj}(0),
\end{cases}
\label{supp:eq:boundary-summary-s0}\\
I_L(\tfrac12,T)&=0\qquad(0\le T\le\infty),\label{supp:eq:boundary-summary-half}\\
I_L(s,0)&=
\begin{cases}
-1,&0\le s<\tfrac12,\\
0,&s=\tfrac12,\\
+1,&\tfrac12<s\le1,
\end{cases}
\label{supp:eq:boundary-summary-T0}\\
I_L(s,\infty)&=0\qquad(0\le s\le1).
\label{supp:eq:boundary-summary-Tinf}
\end{align}
The first line is a thermodynamic-limit statement; the remaining three hold
at every finite $L$.  No value is assigned to the critical point
$(s,T)=(0,T_c^{\rm proj}(0))$.

The full fusion projector instead pins the winding sector. For the all-$+$ projector $P_+$, $S_\Sigma^{(1)}P_+=P_+$ fixes the twisted membrane to $+1$ irrespective of phase; choosing the member of the $2^6$ sector family of Sec.~\ref{supp:sm:fusion} matched to the twisted fixed-point ground state likewise replaces the winding average by $1$. The invariant is therefore defined in $\rho_{\rm loc}$.

The projected-paramagnet endpoint also admits a complementary charge-pair
diagnostic. Relaxing the Gauss projector at two vertices,
$P_v\to\frac{1-S^{(1)}_v}{2}$ at $v$ and $v'$, inserts two endpoint defects.
In the loop representation their correlator is the open-string sum, which
under the map of Sec.~\ref{supp:sm:s1-ising} is exactly the Ising two-point
function $\langle\sigma_v\sigma_{v'}\rangle$. It decays exponentially at low
temperature, where the charges are bound by electric strings of finite
tension, and tends to a nonzero constant at high temperature, where the
strings condense and the charges deconfine. The Ising order parameter is
therefore a disorder operator of the projected paramagnet, while the
specific heat locates the same three-dimensional Ising transition without
characterizing its two phases.

\section{Conditional interior quantization}
\label{supp:sm:interior}

We now extend the fixed-point mechanism to the open regions with
$s<\tfrac12$. The exact reflection in Eq.~(\ref{supp:eq:tw-reflection}) will then
determine the $s>\tfrac12$ half.

\subsection{Conserved membrane sector and two-copy worldsheet expansion}

In the joint eigenbasis of the mutually
commuting stabilizers $\{K_p,K_l\}$, the Hamiltonian is not diagonal at
general $s$, but its matrix elements follow from $X_p=K_pB_p$ and
$X_l=K_lB_l$. On a closed surface $\prod_{p\in\Sigma}B_p=I$, and therefore
\begin{equation}
S_\Sigma^{(1)}=\prod_{p\in\Sigma}K_p .
\end{equation}
Its eigenvalue $m\coloneqq m_\Sigma\in\{\pm1\}$ is the winding parity of the
flipped-$K_p$ loops.
Every local flip changes an even number of plaquettes of $\Sigma$ and therefore
preserves the sector label $m$.

We now derive the corresponding two-copy worldsheet expansion and its sign.
Here ``copy'' labels the two cell-dual worldsheet species, not replicas of the
physical density matrix; copy $A$ carries the conserved membrane sector $m$,
and copy $B$ is its cell-dual partner.
The diagonal part of $H(s)$ is
$-(1-s)(\sum_pk_p+\sum_lk_l)$. In the expansion of $e^{-\beta H(s)}$, an
$X_l$ insertion flips the four $k_p$ adjacent to $l$ and has matrix element
$s\,k_l$, whereas an $X_p$ insertion flips the four $k_l$ around $p$ and
has matrix element $s\,k_p$. The minus signs in the Hamiltonian become
positive coefficients in $-\beta H$; the only remaining signs are the
instantaneous eigenvalues $k_l$ and $k_p$.

For each copy, periodicity of the trace requires every stabilizer eigenvalue
to return to its initial value. The flip events consequently have no
imaginary-time boundary. The cube and vertex constraints remove their
spatial boundaries. Thus the $A$ and $B$ event complexes are closed mod-$2$
two-cycles
\begin{equation}
\mathcal S_A,\mathcal S_B\subset T^3\times S^1 .
\end{equation}
Consider first the untwisted trace. Following a $B$-copy defect through
imaginary time, its eigenvalue contributes $-1$ precisely when its
worldsheet crosses an $A$ event; the converse statement holds after
interchanging the copies. Hence the product of all local eigenvalue
prefactors is the mod-$2$ intersection sign
\begin{equation}
\prod_{e_A}k_l(e_A)\prod_{e_B}k_p(e_B)
=(-1)^{\mathcal S_A\cdot\mathcal S_B}.
\end{equation}
Here $e_A$ and $e_B$ run over the flip-insertion events of copies $A$ and
$B$, respectively.
Absorbing the twist by $k_p\mapsto\eta_pk_p$ leaves the positive magnitudes
unchanged. Each $B$ event then acquires the additional local factor
$\eta_p$, whose product is
\begin{equation}
(-1)^{\mathcal S_B\cdot(\gamma\times S^1)} .
\end{equation}

The K\"unneth decomposition
\[
\begin{split}
H_2(T^3\times S^1;\mathbb Z_2)
\simeq{}& H_2(T^3;\mathbb Z_2)\\[-2pt]
&{}\oplus\big[H_1(T^3;\mathbb Z_2)\otimes H_1(S^1;\mathbb Z_2)\big]
\end{split}
\]
allows us to write
\begin{equation}
[\mathcal S_A]\coloneqq(a_s,a_\tau),\qquad
[\mathcal S_B]\coloneqq(b_s,b_\tau),
\end{equation}
where $a_s,b_s\in H_2(T^3;\mathbb Z_2)$ are purely spatial sheets and
$a_\tau,b_\tau\in H_1(T^3;\mathbb Z_2)$ label cylinders containing the time
circle. The intersection form pairs the two K\"unneth summands,
$\mathcal S_A\cdot\mathcal S_B
=\langle a_s,b_\tau\rangle+\langle b_s,a_\tau\rangle$, while
$\mathcal S_B\cdot(\gamma\times S^1)=\langle b_s,[\gamma]\rangle$.
The total sign is therefore
\begin{equation}\label{supp:eq:homology-sign-sm}
(-1)^I,\qquad
I\coloneqq\langle a_s,b_\tau\rangle+
  \langle b_s,a_\tau+[\gamma]\rangle\pmod 2 .
\end{equation}
It is $+1$ unless at least one copy contains a noncontractible spatial
sheet.

After the eigenvalue signs have been separated, the diagonal Boltzmann
factors and the magnitudes $s$ of all insertions factorize between the two
species. Denote the resulting positive weights by $W_{A,m}$ and $W_B$ on
the worldsheet configuration spaces $\Omega_{A,m}$ and $\Omega_B$.
Restricting the original projected trace to the conserved membrane
eigenvalue $m$ produces exactly the first sum below; hence
$Z_{\rm coupled}^{(m)}$ is the $Z_m$ of the Letter, with no omitted
normalization or topological sector. Dropping only the intersection sign
gives the product of the two positive-copy partition functions:
\begin{align}
Z_{\rm coupled}^{(m)}
 &\coloneqq\sum_{\mathcal S_A\in\Omega_{A,m}}
   \sum_{\mathcal S_B\in\Omega_B}
   W_{A,m}(\mathcal S_A)W_B(\mathcal S_B)(-1)^I,\label{supp:eq:signed-sum}\\
Z_{A,m}Z_B
 &\coloneqq\sum_{\mathcal S_A\in\Omega_{A,m}}
   \sum_{\mathcal S_B\in\Omega_B}
   W_{A,m}(\mathcal S_A)W_B(\mathcal S_B).\label{supp:eq:positive-sum}
\end{align}
Let $Z_{A,m}^{(a_s)}$ and $Z_B^{(b_s)}$ denote the positive partition
functions restricted to the indicated spatial classes. Then
$Z_{A,m}\coloneqq\sum_{a_s}Z_{A,m}^{(a_s)}$ and
$Z_B\coloneqq\sum_{b_s}Z_B^{(b_s)}$. Define
\begin{equation}
\epsilon_{A,m}\coloneqq
\frac{\sum_{a_s\ne0}Z_{A,m}^{(a_s)}}{Z_{A,m}},\qquad
\epsilon_B\coloneqq\frac{\sum_{b_s\ne0}Z_B^{(b_s)}}{Z_B}.
\end{equation}
Since the sign in Eq.~(\ref{supp:eq:signed-sum}) differs from $+1$ only on this
union of events,
\begin{equation}\label{supp:eq:factorization-bound}
\left|\frac{Z_{\rm coupled}^{(m)}}{Z_{A,m}Z_B}-1\right|
\le2(\epsilon_{A,m}+\epsilon_B).
\end{equation}

\subsection{Hypothesis (H) and the proxy bound}

On the portion of the
$s<\tfrac12$ half-plane where the reduction is used, Hypothesis (H) is the
uniform bound
\begin{equation}\label{supp:eq:H-sm}
\begin{aligned}
\epsilon_{A,m}+\epsilon_B&\le Ce^{-\tau_{\rm sp}L^2},\\[-2pt]
0<C<\infty,&\qquad \tau_{\rm sp}>0,\qquad m=\pm1 .
\end{aligned}
\end{equation}
For each fixed $(s,T)$ where the hypothesis is invoked, $C$ and
$\tau_{\rm sp}$ are independent of both $L$ and $m$.  It is an independent
physical input, not an automatic consequence of being
on one side of the zero-temperature self-dual point. Define the complete
positive-copy splitting
\begin{equation}
\Delta F_A\coloneqq\ln(Z_{A,+}/Z_{A,-}).
\end{equation}
Applying Eq.~(\ref{supp:eq:factorization-bound}) in the two sectors gives
\begin{equation}\label{supp:eq:interior-splitting}
F_\Sigma=\Delta F_A+O(e^{-\tau_{\rm sp}L^2})
\end{equation}
and hence
\begin{equation}\label{supp:eq:proxy-bound}
\left|\langle S_\Sigma^{(1)}\rangle_{\rm tw}^{P_{\rm loc}}
-(-1)^{\Sigma\cdot\gamma}\tanh\frac{\Delta F_A}{2}\right|
\le C' e^{-\tau_{\rm sp}L^2}.
\end{equation}
Here $C'<\infty$ is independent of $L$ at the fixed $(s,T)$ under
consideration.

\subsection{Conditional quantization theorem}

Suppose, in addition to (H), that
copy $A$ has a positive winding-line tension below the arm,
\begin{equation}
\Delta F_A\ge\kappa(s,T)L-O(\ln L),\qquad \kappa(s,T)>0,
\end{equation}
and a positive interface tension above it,
\begin{equation}
\begin{aligned}
\left|\frac{Z_{A,+}-Z_{A,-}}{Z_{A,+}+Z_{A,-}}\right|
&\le C_{\rm int} e^{-\sigma(s,T)L^2},\\[-2pt]
0<C_{\rm int}<\infty,&\qquad \sigma(s,T)>0,
\end{aligned}
\end{equation}
with $C_{\rm int}$ independent of $L$.
Equation~(\ref{supp:eq:proxy-bound}) then gives
$\mathcal I=(-1)^{\Sigma\cdot\gamma}=-1$ for odd intersection in the
$s<\tfrac12$ low-temperature phase and $\mathcal I=0$ in the
thermally disordered phase. Exact reflection,
Eq.~(\ref{supp:eq:tw-reflection}), gives $\mathcal I=+1$ in the reflected
$s>\tfrac12$ low-temperature phase and again zero above its arm. The
self-dual wall $s=\tfrac12$ below $T_{\rm proj}^*$ is excluded from the open
phases; its two one-sided limits have opposite signs. This completes the
conditional implication from the three positive tensions to the three
quantized phase values.

\section{Numerical tests of the membrane index}
\label{supp:sm:index-numerics}

\subsection{One-copy membrane proxy and spatial-sheet sector ratios}
\label{supp:sm:stw-num}
We now describe the two numerical calculations that test the assumptions of
Sec.~\ref{supp:sm:interior}: the winding-sector splitting in the trivial spatial
class and the direct ratio between spatial-sheet classes.

\subsubsection{Representation and membrane-sector measurement}

The $A$ copy is
simulated in its electric (worldsheet) representation, with spatial-link
variables $x_l(\tau)$ on $N_\tau$ imaginary-time slices. For the base step
$\Delta\tau_0\coloneqq0.2$, we choose
\begin{equation}
N_\tau\coloneqq\max\{4,\operatorname{round}(\tilde\beta/\Delta\tau_0)\},
\qquad
\Delta\tau_{\rm eff}\coloneqq\tilde\beta/N_\tau ,
\end{equation}
where $\tilde\beta\coloneqq s\beta$ is the inverse temperature of the $A$
copy at unit flip coupling; it is distinct from the unit-$B$ rescaling
$\tilde T$ of Sec.~\ref{supp:sm:tstar-scaling}. Down links form closed defect
loops, the Gauss
constraint is diagonal, and plaquette-flip events
$n_p(\tau)\in\{0,1\}$ enforce
$x_l(\tau)x_l(\tau+1)=(-1)^{\sum_{p\ni l}n_p(\tau)}$.
All weights are positive:
$e^{+\Delta\tau_{\rm eff}h^x_Ax_l(\tau)}$ per link-slice,
with $h^x_A\coloneqq(1-s)/s$, and $\tanh(\Delta\tau_{\rm eff})$ per event.
The local moves
(paired temporal events, spatial-cube events, and plaquette columns) preserve
both the winding parity $m$ and the purely spatial class $a_s$.  Within the
fixed $(m,a_s=0)$ class used for the map, they are ergodic.  The two winding
sectors $m=+$ (even) and $m=-$ (odd) are simulated independently, the latter
seeded with a single winding line.

Thermodynamic integration in a field strength $\lambda$ deforms the
link-slice weight to
$e^{+\lambda\Delta\tau_{\rm eff}h^x_Ax_l(\tau)}$ and measures
\begin{equation}\label{supp:eq:stw-ti}
\Delta F_A^{(0)}
=-\Delta\tau_{\rm eff}\,h^x_A\!\int_0^1\!d\lambda\,
\big[\langle{\textstyle\sum}x\rangle_{-,0}
-\langle{\textstyle\sum}x\rangle_{+,0}\big]_\lambda ,
\end{equation}
where
\begin{equation}
\Delta F_A^{(0)}
\coloneqq\ln\frac{Z_{A,+}^{(a_s=0)}}{Z_{A,-}^{(a_s=0)}} .
\end{equation}
Here $\sum x\coloneqq\sum_{l,\tau}x_l(\tau)$, and the subscripts
$(\pm,0)$ specify membrane parity $m=\pm1$ and the trivial spatial-sheet
class $a_s=0$.
The anchor at $\lambda=0$ is exact because flipping one winding line is a
weight-preserving bijection.  Equation~(\ref{supp:eq:stw-ti}) does not sum over
the omitted spatial classes.  Under (H),
\begin{equation}\label{supp:eq:restricted-to-full}
\begin{aligned}
Z_{A,m}&=Z_{A,m}^{(0)}
\big[1+O(e^{-\tau_{\rm sp}L^2})\big],
\\
\Delta F_A&=\Delta F_A^{(0)}
+O(e^{-\tau_{\rm sp}L^2}).
\end{aligned}
\end{equation}
Only after this step and the two-copy bound above does the proxy
$(-1)^{\Sigma\cdot\gamma}\tanh(\Delta F_A^{(0)}/2)$ approximate the physical
twisted membrane.

\subsubsection{Direct spatial-sheet sector ratio}

We test the omitted step with a
second calculation that forces one spatial event sheet.  For
$a\ne0$, define
\begin{equation}\label{supp:eq:Fsp-sm}
\begin{aligned}
F_{\rm sp}^{A,m}(a;L)
&\coloneqq-\ln[Z_{A,m}^{(a_s=a)}/Z_{A,m}^{(a_s=0)}],\\
\tau_{A,m}(a;L)&\coloneqq L^{-2}F_{\rm sp}^{A,m}(a;L).
\end{aligned}
\end{equation}
Let $S$ be a noncontractible plane of $L^2$ plaquette events representing
$a$, let $N(n)$ be the total event number, and let $N_S$ be the number of
occupied events on $S$.  The factor $W_{\rm rest}(n)$ denotes all positive
weights and constraints that are unchanged by the interpolation.  Let
$\kappa_p\coloneqq-\ln\tanh(\Delta\tau_{\rm eff})$ be the physical event
penalty. The
constraint-preserving bijection $n\mapsto n\oplus S$ allows both sectors to
be written over the reference configuration space $\mathcal C_0$:
\begin{equation}\label{supp:eq:sheet-interpolation-sm}
\begin{split}
Z_\lambda\coloneqq\sum_{n\in\mathcal C_0}W_{\rm rest}(n)
\exp\!\big\{-\kappa_p[&(1-\lambda)N(n)\\[-2pt]
&+\lambda N(n\oplus S)]\big\}.
\end{split}
\end{equation}
Thus $Z_0=Z_{A,m}^{(0)}$, $Z_1=Z_{A,m}^{(a)}$, and
\begin{equation}\label{supp:eq:sheet-ti-sm}
\begin{aligned}
F_{\rm sp}^{A,m}(a;L)
&=\kappa_p\int_0^1d\lambda\,\langle Q\rangle_\lambda,\\
Q&\coloneqq N(n\oplus S)-N(n)=L^2-2N_S .
\end{aligned}
\end{equation}
The constant $\lambda\kappa_pL^2$ contained in $Q$ is part of the
normalization; omitting it would compute the wrong sector ratio.

At $(s,T)=(0.30,0.30)$ and $\Delta\tau_{\rm eff}=0.2$, a fixed-$\lambda$
integration and an independent Wang--Landau simulated-tempering calculation
agree at the percent level.  Frozen-weight production estimates of
$\tau_{A,+}$ are $1.512,1.519,1.534,1.526$ for
$L=6,8,10,12$, respectively.  At $L=8$, direct integrations for the
$xy,xz,yz$ generator sheets give $1.5210(19)$, $1.5186(34)$, and
$1.5126(45)$, statistically consistent with cubic symmetry.  The same
finite-$\Delta\tau_{\rm eff}$ estimator remains positive near the self-dual line,
where it gives approximately $0.71,0.72,0.77,0.76$ for
$(s,T)=(0.48,0.30)$ and $L=6,8,10,12$, also at
$\Delta\tau_{\rm eff}=0.2$. At the representative high-temperature point
$(s,T)=(0.30,1.20)$, it is approximately $1.60$ with
$\Delta\tau_{\rm eff}=0.0625$.
All values in this paragraph use the trivial temporal class $a_\tau=0$;
for the chosen membrane this is the even sector $m=+1$.  The odd sector was
not measured in this sheet-cost calculation.

In the multicanonical calculation $\lambda$ is dynamical during adaptation;
the weights are then frozen and the endpoint histogram ratio supplies the
residual reweighting correction. Opposite endpoint initializations agree
within the observed seed spread. The finite modification factor, limited
round trips, and common $\Delta\tau_{\rm eff}=0.2$ in these two reference
calculations make this evidence for a positive area cost rather than a
precision continuum extrapolation.

Copy $B$ is related to copy $A$ by cubic cell duality.  Moreover, changing
the winding background can be implemented by a logical loop that changes
only $O(L)$ local terms.  The partition-function Lipschitz bound at fixed
$\beta$ then gives
\begin{equation}
\big|F_{\rm sp}^{A,+}(a;L)
-F_{\rm sp}^{A,-}(a;L)\big|=O(L),
\end{equation}
which makes the leading $L^2$ rate common to the two backgrounds; the same
conclusion holds for copy $B$.  Cubic symmetry reduces the seven nonzero
classes to representatives of Hamming weights one, two, and three.  The
calculation above tests the weight-one orbit.  The $(1,1,0)$ and
$(1,1,1)$ classes, a continuum extrapolation, and uniform coverage of both
open regions remain untested.  The data therefore support one constituent
of (H), but do not establish the full hypothesis.

\subsubsection{Validation and domain}

The estimator passes the exact $s\to0$
loop-gas anchor of Secs.~\ref{supp:sm:s1-ising}--\ref{supp:sm:s1-twist}, where the same
observable is computed independently by extended-ensemble Ising Monte Carlo.
Low-temperature interior points approach the expected quantized values,
while points above the dome approach zero when the discretization is taken
with $\Delta\tau_{\rm eff}=\tilde\beta/N_\tau$ exactly; these are numerical consistency
checks, not additional exact anchors.  We do not use the decoupled-$A$ proxy
as a direct physical estimator for $s>\tfrac12$, where the spatial-sheet
hypothesis need not hold.  The physical membrane on that half follows from
the exact twisted-reflection identity, Eq.~(\ref{supp:eq:tw-reflection}), and
retains the conditional status of the measured half.

The membrane-map updates preserve both $m$ and $a_s$, and hence do
not themselves test (H).  Equations~(\ref{supp:eq:Fsp-sm})--(\ref{supp:eq:sheet-ti-sm})
provide a separate direct test of one spatial-class ratio.  The distinction
between these two numerical calculations is essential.

Figure~1(b) of the Letter shows this conditional one-copy proxy at $L=8$
with base step $\Delta\tau_0=0.2$.  The directly measured quantity is
$\Delta F_A^{(0)}$: it approaches opposite plateaus in the two
low-temperature regions and is statistically consistent with zero above the
independently estimated arm $T_c^{\rm proj}(s)$.  Interpreting those values as
the physical thermodynamic index requires Hypothesis (H) and the one-copy
tension assumptions of Sec.~\ref{supp:sm:interior}.  At the exact endpoints
$s=0,1$, the index instead follows directly from the loop gases of
Secs.~\ref{supp:sm:s1-loops} and \ref{supp:sm:s1-twist}, without thermodynamic
integration or Hypothesis (H).

\subsection{Hypothesis-independent finite-size evaluation of the membrane index}
\label{supp:sm:hfree-qmc}

This calculation is distinct from the conditional one-copy color map in
Fig.~1(b) of the Letter. It retains every spatial-sheet and temporal-winding
sector of two independently sampled positive worldsheet species and performs
the signed sum exactly at finite $L$. It therefore assumes neither
spatial-sheet suppression nor equality of the two species. The sampling and
acceptance rules were fixed before the physical free energies and $I_L$ were
examined.

\subsubsection{Sector tables and exact reconstruction}

Here copies $A$ and $B$ are the two positive worldsheet species introduced in
Sec.~\ref{supp:sm:interior}, not replicas of the density matrix. Copy $A$ carries
the membrane-sector constraint, while copy $B$ is its cell-dual partner. For
$C=A,B$, let $c_s\in H_2(T^3;\mathbb Z_2)$ be the homology class of its
spatial sheets, let $c_t\in H_1(T^3;\mathbb Z_2)$ be the class of its temporal
winding lines, and let $Z_C(c_s,c_t)>0$ be the corresponding positive
partition sum. We write
\begin{equation}
\ell_C(c_s,c_t)\coloneqq\ln Z_C(c_s,c_t).
\end{equation}
The same mod-$2$ intersection pairing denoted by
$\Sigma\mathbin{\cdot}\gamma$ above is written
$\langle\ ,\ \rangle:H_2\times H_1\to\mathbb Z_2$ when its arguments are
homology-coordinate vectors. Explicitly, if
$p=(p_{xy},p_{xz},p_{yz})$ labels sheet classes and
$q=(q_x,q_y,q_z)$ labels winding lines, then
\begin{equation}
\langle p,q\rangle
=p_{xy}q_z+p_{xz}q_y+p_{yz}q_x\pmod 2.
\label{supp:eq:hfree-pairing}
\end{equation}
The measured membrane and inserted twist obey $[\Sigma]=[xy]$ and
$[\gamma]=[z]$, respectively. Hence
$\Sigma\mathbin{\cdot}\gamma
=\langle[\Sigma],[\gamma]\rangle=1$ by
Eq.~(\ref{supp:eq:hfree-pairing}).

Each positive $8\times8$ sector table is determined relative to
$Z_C(0,0)$ by the temporal ratios and spatial-sheet costs
\begin{equation}
\begin{aligned}
r_C(c_t)&\coloneqq
 \ln\frac{Z_C(0,c_t)}{Z_C(0,0)},\\
F_{\rm sp}^{C}(c_s,c_t)&\coloneqq
 -\ln\frac{Z_C(c_s,c_t)}{Z_C(0,c_t)},\\
\ell_C(c_s,c_t)-\ell_C(0,0)&=r_C(c_t)-F_{\rm sp}^{C}(c_s,c_t).
\end{aligned}
\label{supp:eq:hfree-positive-table}
\end{equation}
Cubic-axis permutations reduce the 64 entries to 20 exact orbits: one fixes
the arbitrary normalization, three are nontrivial temporal orbits, and 16
have $c_s\ne0$. Thus 19 measured ratios determine each table, with no
identification of $A$ and $B$.

To reconstruct the physical trace, use copy-specific labels
$a_s,b_s\in H_2(T^3;\mathbb Z_2)$ and
$a_\tau,b_\tau\in H_1(T^3;\mathbb Z_2)$, consistently with the K\"unneth
labels in Sec.~\ref{supp:sm:interior}, and define
\begin{equation}
\begin{aligned}
\Omega_m&\coloneqq
\{(a_s,a_\tau,b_s,b_\tau):
(-1)^{\langle[\Sigma],a_\tau\rangle}=m\},\\
\chi(a_s,a_\tau,b_s,b_\tau)&\coloneqq
(-1)^{\langle a_s,b_\tau\rangle+
\langle b_s,a_\tau+[\gamma]\rangle}.
\end{aligned}
\end{equation}
The physical partition sum in membrane sector $m=\pm1$ is then
\begin{equation}
Z_m^{\rm phys}\coloneqq\sum_{\Omega_m}\chi(a_s,a_\tau,b_s,b_\tau)
e^{\ell_A(a_s,a_\tau)+\ell_B(b_s,b_\tau)}.
\label{supp:eq:hfree-signed-reconstruction}
\end{equation}
Each value of $m$ contains 2048 terms. Positive and negative terms are summed
separately in log space before their difference is taken. The calculation
requires $Z_m^{\rm phys}$ to be positive and finite and monitors the resolved
fraction of that cancellation,
\begin{equation}
{\cal C}_m\coloneqq
\frac{\left|\sum_{\Omega_m}\chi(a_s,a_\tau,b_s,b_\tau)
e^{\ell_A(a_s,a_\tau)+\ell_B(b_s,b_\tau)}\right|}
{\sum_{\Omega_m}e^{\ell_A(a_s,a_\tau)+\ell_B(b_s,b_\tau)}}.
\label{supp:eq:hfree-cancellation}
\end{equation}
For each sign, let ${\cal C}_m^*$ denote a bootstrap replicate of
Eq.~(\ref{supp:eq:hfree-cancellation}).  The one-sided 99\% bootstrap-basic lower bound
$\max[0,2{\cal C}_m-Q_{0.99}({\cal C}_m^{*})]$ must exceed $10^{-4}$, where
$Q_{0.99}$ is the linearly interpolated bootstrap quantile. Finally,
\begin{equation}
I_L=(-1)^{\langle[\Sigma],[\gamma]\rangle}
\tanh\!\left[
\frac{\ln Z_+^{\rm phys}-\ln Z_-^{\rm phys}}{2}
\right].
\label{supp:eq:hfree-index}
\end{equation}
Equations~(\ref{supp:eq:hfree-signed-reconstruction}) and
(\ref{supp:eq:hfree-index}) are exact at the simulated $L$ and imaginary-time
discretization; unlike Eq.~(\ref{supp:eq:restricted-to-full}), they omit no
homology class.

\subsubsection{Worldsheet variables and updates}

The positive tables are sampled on an $L^3\times N_\tau$ spacetime lattice,
where $N_\tau$ is the number of imaginary-time slices.  For each
$C\in\{A,B\}$, let $x_{C,l}(\tau)=\pm1$ be its directed-link worldline
variable and let $n_{C,p}(\tau)\in\{0,1\}$ record a plaquette-flip event
between adjacent slices.  Write
$N(n_C)\coloneqq\sum_{p,\tau}n_{C,p}(\tau)$.  Copy $B$ lives on the
cell-dual cubic lattice: its links and plaquettes are dual to those of copy
$A$.  Cell duality gives both copies the same positive local action at the
link--plaquette-symmetric coupling used here, but their variables, random
streams, and sector tables are sampled independently. With
$\tilde\beta=s/T$, $h_{\rm ws}^x\coloneqq(1-s)/s$, and
$\Delta\tau_{\rm eff}=\tilde\beta/N_\tau$, set
$\phi\coloneqq\Delta\tau_{\rm eff}h_{\rm ws}^x$ and
$\kappa_{\rm evt}\coloneqq-\ln\tanh(\Delta\tau_{\rm eff})$. The weight for either
copy is, up to sector-independent constants,
\begin{equation}
W_C[x_C,n_C]\propto
\exp\!\left[\phi\sum_{l,\tau}x_{C,l}(\tau)-\kappa_{\rm evt}N(n_C)\right],
\label{supp:eq:hfree-weight}
\end{equation}
subject, for each copy, to the Gauss constraint and closure of the event worldsheet in
spacetime. Paired temporal-event moves, cube-boundary moves, and
plaquette-column moves change $x_C$ and $n_C$ locally while preserving both
constraints.

The free-energy ratios in Eq.~(\ref{supp:eq:hfree-positive-table}) are obtained by
an extended-ensemble bridge between the two endpoint homology classes. A
fixed winding line or sheet would select only one translated representative,
so its support is itself sampled as an auxiliary coordinate. For a temporal
class $q_t$, the set and number of translated line supports are
\begin{equation}
{\cal R}_{q_t}\coloneqq\prod_{d:(q_t)_d=1}
(\mathbb Z_L\times\mathbb Z_L),
\qquad |{\cal R}_{q_t}|=L^{2|q_t|},
\end{equation}
where $|q_t|$ is the number of nonzero components. Each selected winding
line is repeated on every imaginary-time slice, so only its two transverse
spatial coordinates label translations. For a spatial class
$q_s$, they are
\begin{equation}
{\cal R}_{q_s}\coloneqq\prod_{o:(q_s)_o=1}
(\mathbb Z_L\times\mathbb Z_{N_\tau}),
\qquad |{\cal R}_{q_s}|=(LN_\tau)^{|q_s|},
\end{equation}
where $|q_s|$ likewise counts the nonzero components of $q_s$.

For a class $q$, let $n_\lambda\coloneqq25,49,73$ when $|q|=1,2,3$, respectively,
let $J\coloneqq n_\lambda-1$, and use the reflection-symmetric grid
$\lambda_j\coloneqq j/J$.  Set $A_0\coloneqq-\ln W_C$, with its irrelevant common
additive constant understood.  For a translated spatial
support $S_r$, define
$N_{S_r}\coloneqq\sum_{(p,\tau)\in S_r}n_{C,p}(\tau)$ and
\begin{equation}
Q_r\coloneqq |S_r|-2N_{S_r},\qquad
A^{\rm sp}_{\lambda,r}\coloneqq A_0+\kappa_{\rm evt}\lambda Q_r .
\label{supp:eq:hfree-spatial-bridge-action}
\end{equation}
The constant $\kappa_{\rm evt}\lambda|S_r|$ is included. For a translated temporal
support $r$, let $X_r\coloneqq\sum_{(l,\tau)\in r}x_{C,l}(\tau)$ and use
\begin{equation}
A^{\rm line}_{\lambda,r}\coloneqq A_0+2\phi\lambda X_r .
\label{supp:eq:hfree-line-bridge-action}
\end{equation}
Orbit bin $j$ combines the trivial-class branch at $\lambda_j$ with the
target-class branch at $1-\lambda_j$, both with support $r$.  With
$A=A^{\rm sp}$ or $A^{\rm line}$, frozen sampled weight
$\exp[-A_{\lambda,r}-b_j]$, and production visit count $H_j$, the endpoint
estimators are
\begin{equation}
\begin{aligned}
F_{\rm sp}^{C}(c_s,c_t)
 &=b_0-b_J+\ln(H_0/H_J),\\
r_C(c_t)&=-b_0+b_J-\ln(H_0/H_J).
\end{aligned}
\label{supp:eq:hfree-bridge-estimators}
\end{equation}
Both endpoints contain the same two reflected branches and the same support
multiplicity, so this auxiliary factor cancels and no entropic correction is
inserted.

Along with the local moves, the bridge proposes neighboring bins and
translated supports and uses an involution that exchanges the endpoint
classes while sending $\lambda$ to $1-\lambda$.  For any symmetric proposal,
the acceptance probability is $\min(1,\alpha)$, where
\begin{equation}
\ln\alpha\coloneqq-[A_{\lambda',r'}-A_{\lambda,r}]-(b_{j'}-b_j),
\label{supp:eq:hfree-metropolis}
\end{equation}
which enforces detailed balance; the reflected support flip is
rejection-free because it leaves $A$ invariant.  A separate exact
Metropolis proposal directly exchanges physical endpoints and is recorded
only as a mixing diagnostic.  Exhaustive $L=2$ tests covered every nonzero
class, support, reflected branch, and directed bin edge, including a path
between translated representatives. At $L=4$, the action changes and
Gauss/closure identities were checked for every move type, and every
accepted chain completed at least one adjacent-path round trip. These are
validation and mixing checks at the simulated size, not a proof of
ergodicity for arbitrary $L$.

\subsubsection{Trotter refinement and sampling inventory}

The time discretization is chosen by
\begin{equation}
\begin{aligned}
N_\tau^{(0)}&\coloneqq\max\!\left[4,
\operatorname{round}\!\left(\frac{s/T}{\Delta\tau_{\rm target}}\right)
\right],\\[-2pt]
N_\tau&\coloneqq fN_\tau^{(0)},&
\Delta\tau_{\rm eff}&\coloneqq\frac{s/T}{N_\tau},
\end{aligned}
\label{supp:eq:hfree-trotter}
\end{equation}
with $\Delta\tau_{\rm target}=0.2$ and $f=1,2,4$. Thus $N_\tau$ and
$\Delta\tau_{\rm eff}$ are fixed before a chain is run. The refinement factor
$f$ is not a physical coupling: it labels successively finer approximations
to imaginary time. Results at different $f$ are consequently convergence
checks, not independent measurements of a common finite-step observable.
Averaging them would retain an uncontrolled Trotter bias, so they are
reported separately; no continuum extrapolation is made.

For each copy, the 19 nontrivial orbit ratios are sampled in four independent
waves. Two waves start at each endpoint, in the fixed order $[0,1,0,1]$.
Every $(s,T,f)$ cell therefore contains
\begin{equation}
19\ \text{ratios}\times2\ \text{copies}\times4\ \text{waves}=152
\ \text{independent chains}.
\end{equation}
Each accepted chain contains $8{,}000{,}000$ production sweeps. The complete
inventory is required: no chain is dropped, imputed, or replaced by a
different random seed.

\subsubsection{Calibration, acceptance, and uncertainty}

Calibration uses only bridge-bin counts, not a physical free energy or
$I_L$.  In round 0 the offsets $b_j$ learned during adaptation are frozen,
and the round passes only if every bin is visited.  If its counts are $H_j$,
it proposes the prescribed centered half-step
\begin{equation}
\delta b_j\coloneqq\frac12\left[
\ln H_j-\frac1{n_\lambda}\sum_{k=0}^{J}\ln H_k
\right].
\label{supp:eq:hfree-bias-update}
\end{equation}
Round 1 samples with exactly $b_j+\delta b_j$ and must again visit every
bin; it makes no further update.  Production is permitted only under this
set of offsets covered in this second frozen round.  An uncovered round, a
return to an earlier set of offsets, or use of an unvalidated offset rejects
the entire chain in this representative campaign.

For reproducibility, let $u=1,\ldots,19$ label a measured orbit ratio, let
$\widehat q_{wu}$ be that ratio from wave $w$, and let $V^{(r)}_{wu}$ be its
delete-one-contiguous-block jackknife variance
after rebinning the 32 production blocks by $r=1,2,4$. The within-wave
variance is
\begin{equation}
v_{wu}\coloneqq\max_{r\in\{1,2,4\}}V^{(r)}_{wu}
+\max\!\left[0,
\frac{d_{wu}^2-v^{\rm early}_{wu}-v^{\rm late}_{wu}}{4}
\right],
\label{supp:eq:hfree-chain-variance}
\end{equation}
where $d_{wu}$ is the difference between the estimates from the second and
first halves, and $v^{\rm early/late}_{wu}$ are their multiscale jackknife
variances. A Welch--Satterthwaite test records the corresponding stationarity
diagnostic. For four waves, define
$\overline q_u\coloneqq\tfrac14\sum_w\widehat q_{wu}$,
$S_u^2\coloneqq\tfrac13\sum_w(\widehat q_{wu}-\overline q_u)^2$, and
$\overline v_u\coloneqq\tfrac14\sum_wv_{wu}$. The diagonal random-effects rule is
\begin{equation}
\tau_u^2\coloneqq\max(0,S_u^2-\overline v_u),\qquad
\operatorname{Var}(\overline q_u)\coloneqq
\frac1{16}\sum_wv_{wu}+\frac{\tau_u^2}{4}.
\label{supp:eq:hfree-random-effects}
\end{equation}

Equation~(\ref{supp:eq:hfree-signed-reconstruction}) is propagated through 5000
Gaussian parametric-bootstrap draws from these diagonal covariances, using
independent random streams for $A$ and $B$; at least 99\% of the draws must
produce finite positive partition sums. A failed or nonfinite draw is assigned
${\cal C}_m^*=0$ in the cancellation bound, so it cannot make that gate easier
to pass. In addition, waves $(0,1)$ and
$(2,3)$ form two balanced-start tables for each copy. Crossing the two
$A$ tables with the two $B$ tables gives four reconstructed indices. If
$s_I$ is the bootstrap standard deviation and $R_I$ is half the range of
these four indices, the reported uncertainty scale is
${\cal U}_I\coloneqq\max(s_I,R_I)$. The reported interval is
\begin{equation}
\begin{aligned}
\mathrm{CI}_{95\%}&\coloneqq
\left[I_L\mathbin{\pm}t_{0.975,3}{\cal U}_I\right]\cap[-1,1],\\
t_{0.975,3}&=3.1824463 .
\end{aligned}
\label{supp:eq:hfree-ci}
\end{equation}
The Student-$t$ coefficient uses the three degrees of freedom supplied by
the four independent waves. Because ${\cal U}_I$ also includes the
balanced-start spread, it is an uncertainty scale rather than a standard
error.

Limited sampling efficiency was reported but did not by itself reject a
chain, provided that the chain made at least one complete adjacent-path round
trip. A cell was rejected if its 152-chain inventory was incomplete, an
estimator or uncertainty was nonfinite, calibration failed, or an action,
homology, Gauss, or closure identity failed. The three refinement factors
were judged separately.

\subsubsection{Representative results and limitations}

The five accepted central estimates are those in Table~I of the Letter. At
$(s,T)=(0.30,0.30)$, the three refinement factors give
\begin{equation}
\begin{aligned}
I_4^{(f=1)}&=-0.9999995712,\\[-2pt]
I_4^{(f=2)}&=-0.9999994446,\\[-2pt]
I_4^{(f=4)}&=-0.9999994161.
\end{aligned}
\end{equation}
Their 95\% interval half-widths are, respectively,
$2.0\times10^{-8}$, $3.8\times10^{-8}$, and $3.2\times10^{-8}$. At
$(s,T)=(0.30,1.20)$,
\begin{equation}
\bigl(I_4^{(f=1)},I_4^{(f=2)}\bigr)=(-0.0047491,0.0029787),
\end{equation}
with 95\% intervals $[-0.0199976,0.0104994]$ and
$[-0.0239646,0.0299221]$. All five accepted cells satisfy the hard
requirements above and carry no sampling warning.

The high-temperature $f=4$ calculation does not yield an estimate because
two of its 152 chains failed calibration.  Both were independent waves of
the same most demanding copy-$A$ spatial bridge,
$c_s=c_t=(1,1,1)$, with 73 bins.  One stopped during calibration before
production; the other did not obtain offsets having both complete bin
coverage and final validation.  The entire cell was therefore rejected and
no physical $I_4$ was formed.  This is a sampling/calibration limitation,
not a measured anomalous value.

The additional $(0.70,0.30)$ control runs were not used because their
combined control estimate could not be validated. An independent
reconstruction reproduced all five
accepted estimates and confirmed finite physical partition sums and
uncertainties. Every accepted cell retained its complete 152-chain
inventory; the high-temperature $f=4$ cell was rejected as a whole.

The direct calculation supports $I_4\simeq-1$ and a value consistent with
zero at the two representative points. It is not a phase map, a
$\Delta\tau_{\rm eff}\to0$ extrapolation, or a thermodynamic-limit proof.
Figure~1(b) of the Letter remains the conditional one-copy proxy described in
Sec.~\ref{supp:sm:stw-num}.

\end{document}